\documentstyle[11pt]{article}

\newfont{\msbm}{msbm10}

\newfont{\msbms}{msbm6}  

\def\cst{$C^{\ast}$}

\def\U{{\cal U}}

\def\V{{\cal V}}

\def\Q{{\cal Q}}

\def\Qh{{\cal Q}_{\hbar}}
\def\P{{\cal P}}
\def\cinf{C^{\infty}}
\def\cinfp{\cinf(\P)}
\def\A{{\cal A}}
\def\S{{\cal S}}
\def\ap{\A(\P)}

\def\N{\hbox{\msbm N}}

\def\Z{\hbox{\msbm Z}}

\def\R{\hbox{\msbm R}}

\def\C{\hbox{\msbm C}}

\def\H{{\cal H}}

\def\p{\hbar}
\def\f{\phi}
\def\F{\Phi}
\def\vf{\varphi}

\def\a{\alpha}
\def\b{\beta}
\def\s{\sigma}

\def\ft{{\tilde f}}

\def\h2{${\rm h}(2)$}

\def\pp{\{\, ,\, \}}

\newtheorem{prop}{Proposition}
\newtheorem{cor}{Corol\'ario}
\newtheorem{theo}{Theorem}
\newtheorem{deff}{Definition}

\def\bc{\begin{cor}}
\def\ec{\end{cor}}

\def\bt{\begin{theo}}
\def\et{\end{theo}}

\def\bd{\begin{deff}}
\def\ed{\end{deff}}

\def\bp{\begin{prop}}
\def\ep{\end{prop}}

\def\ba{\begin{eqnarray}}
\def\ea{\end{eqnarray}}

\def\be{\begin{equation}}
\def\ee{\end{equation}}

\def\gg{groupoid}
\def\cc{category}

\begin{document}

\title{An introduction to strict quantization}

\author{J. M. Velhinho}
\date{Faculdade de Ci\^encias,  Universidade da Beira Interior \\ 
R. Marqu\^es D'\'Avila e Bolama, 6201-001 Covilh\~a, Portugal\\
E-mail: jvelhi@ubi.pt}

\maketitle

\begin{abstract}
\noindent
We present a short 
review
of the approach to quantization 
known as  {\em strict (deformation) quantization}, which can be seen as a generalization of the Weyl-Moyal quantization. We include examples and comments
on the process of quantization.


\end{abstract}

\section{Introduction}
\label{intro}
This brief review is my modest contribution to a field and a set of ideias that strongly
influenced my view on the process of {\em quantization}. 
I hope that this introduction to the subject of {\em strict  quantization} and the examples in it can be of help,
and motivate both theoretical and mathematical physicists to this beautiful and highly developed field.
My personal impression is that there is still a whole lot to explore in this area, in particular
concerning extensions of the formalism to the realm of infinite dimensional degrees of freedom.
In fact, it seems clear that quantum field theory could benefit greatly from a strict quantization
approach. Concrete steps have been taken in this direction (see e.g.~\cite{HRS,W,FR} and also \cite{ST}
for a recent application in quantum gravity),
in what is for sure a rather promising area in mathematical physics.

This brief survey focus on finite dimensional examples exclusively, and is mostly inspired
by (parts of) two beautiful books, namely Folland's {\em Harmonic Analysis on Phase Space} \cite{Fo} and Landsman's
synthesis {\em Mathematical Topics Between Classical and Quantum Mechanics} \cite{Lan}.
Another major source of inspiration was V\'arilly's book \cite{Va}.
Concerning broader fields, such as Poisson geometry, groupoids or deformation of algebra structures, we follow also \cite{SW,RL}, as well as \cite{Co}.
An effort is made, hopefully sucessful, to keep the discussion
short and as pedagogical as possible, just enough to give the reader a grasp
of what strict quantization is all about and to motivate him/her to this field.

This work is organized as follows. In the remaining of the present section we give a very brief introduction to the problem of quantization. 
In section \ref{class} we review the formalism of classical mechanics and prepare for the strict quantization approach.
In section \ref{WM} we review the  Weyl-Moyal quantization, which is the prototype of a strict
 deformation quantization. In section \ref{strict} we present what is essentially Landsman's  definition of strict  quantization.
 In the following sections we present methods for the construction of strict quantizations which,
in one way or another, generalize the Weyl-Moyal quantization process. First, in section \ref{gru}
we review the notions of smooth groupoids and associated convolution algebras. Then Connes' tangent groupoid and Landsman's strict quantization of
cotangent bundles of  Riemannian manifolds are discussed, in section \ref{tg}.  Finally, 
we discuss the interesting case of the 2-torus, treated both by 
geometric quantization related and by strict quantization methods.

\bigskip

The formalism of classical mechanics is based on a phase space (the states)
and on functions on that space (the observables), on the set of which two operations
are defined: the product of functions and the Poisson bracket. Quantum mechanics
presents a similar structure, keeping the duality states-observables.
In fact, in quantum mechanics one deals with operators (observables) in a Hilbert space
(the states). The operation that combines observables is now the operator product.
One can extend further the analogy with the classical structure, by means of two
new operations, obtained by symmetrization and antisymetrization of the operator product. 
The anticommutator $\circ $, or Jordan product, is given by
\be
\label{1.0.7}
A\circ B:={1 \over 2}(AB+BA)\ . 
\ee
Following Dirac \cite{Dir}, one defines also the quantum Lie bracket, or quantum
Poisson bracket $[\, ,\, ]_{\hbar }$  by
\be
\label{1.0.8}
[A,B]_{\hbar }:=(AB-BA)/i\hbar \ ,
\ee
where $\hbar =h/2\pi $, with $h$ being Planck's constant. Of course, we have
\be
\label{1.0.9}
AB=A\circ B+{i\hbar \over 2}[A,B]_{\hbar }\ .
\ee
It is tempting to interpret the Jordan product as the quantum equivalent of
the product of functions, together with Dirac's quantum condition stating that 
the quantum  bracket is to be seen as the quantum correspondent of the Poisson bracket.
In broad terms, the transition from a classical description to a quantum description
of a given system - {\it quantization} - is obtained by means of a map $\Q $
from functions on phase space to operators on a Hilbert space.
In this process, one aims at preserving the relations between observables
as much as possible. 
It would therefore be natural to search for maps $\Q $ such that
\be
\label{1.0.10}
\Q (fg)=\Q (f)\circ \Q (g)\, ,
\ee
\be
\label{1.0.11}
\Q (\{f,g\})=[\Q (f),\Q (g)]_{\hbar }\ .
\ee
However, it is well known that such maps cannot be found.
Even Dirac's condition (\ref{1.0.11}) (supplemented with natural requirements, see below)
 is impossible to fullfil exactly, except
for a  restricted set of observables (and typically on cotangent spaces only).
Conditions (\ref{1.0.10}) and (\ref{1.0.11}) should therefore be replaced
by weaker ones, still keeping the purpose  of rigorously implementing  the physical
requirement embodied by those relations. The straightforward {\em canonical quantization}
approach, launched by Dirac, consists in totally relaxing condition (\ref{1.0.10})
and implementing condition  (\ref{1.0.11}) exactly on a small, but sufficiently large,
subalgebra of obervables. Further crucial requirements must be added in this approach,
such as irreducibility (see e.g.~\cite{Go} for a general discussion of  canonical quantization). 

The {\it strict 
quantization} approach \cite{Lan,Ri1} proposes
an asymptotic implementation of conditions (\ref{1.0.10}) and (\ref{1.0.11}).
It involves not a single map $\Q$, but  a family of maps, labeled by a parameter
which we will call $\hbar$, although this is now a free parameter and no longer 
the value of the physical constant.
One then considers a family of maps  $\Q _{\hbar }$, satisfying relations of the type
\be
\label{1.0.12}
\lim_{\hbar \to 0}\Bigl(\Q _{\hbar }(fg)-\Q _{\hbar }(f)\circ 
\Q _{\hbar }(g)\Bigr)=0\ ,
\ee
\be
\label{1.0.13}
\lim_{\hbar \to 0}\Bigl(\Q _{\hbar }(\{f,g\})-[\Q _{\hbar }(f)
,\Q _{\hbar }(g)]_{\hbar}\Bigr)=0\ .
\ee
Let us admit further that each of the $\Q _{\hbar }$ is injective and that,
as a function of $\hbar$, $\Q _{\hbar }$ and $\Q _{\hbar }^{-1}$ are continuous.
In this case,  the operator product induces a family of associative
operations on the functions in phase space:
\be
\label{1.0.14}
f\ast _{\hbar }g:=\Q _{\hbar}^{-1}\Bigl(\Q _{\hbar}(f)\Q _{\hbar}(g)\Bigr)\ ,
\ee
which is a deformation of the standard product, in the sense that 
\be
\lim_{\hbar \to 0}f\ast _{\hbar }g
 = fg \ , \ \ \ \lim_{\hbar \to 0}(f\ast _{\hbar }g-g\ast _{\hbar }f)/i\hbar=\{f,g\}\ .
\ee

\section{Classical systems}
\label{class}
Given a $\cinf$ manifold $\P$, let $\cinf(\P)$ denote
the associative algebra of $\cinf$ complex functions on $\P$,
equipped with the usual product and involution given by
complex conjugation, $f\mapsto \bar f$.
\bd
\label{def1.1.1}
A Poisson structure on a manifold $\P$ is a bilinear operation
$\pp$ 
on $\cinfp$ such that:
\begin{itemize}

\item[{\rm (}i\/{\rm )}] $(\cinf(\P),\pp)$ is a Lie algebra.

\item[{\rm (}ii\/{\rm )}] $\{ f,\cdot\}$ is a derivation on
$\cinfp$ for every $f\in\cinfp$.

\item[{\rm (}iii\/{\rm )}] $\{\bar f,\bar g\}=\overline{\{f,g\}}$.

\end{itemize} 
A manifold with a Poisson structure is said to be a Poisson manifold.
\ed
Given local coordinates $(x_1,\ldots,x_n)$ in $\P$, the Poisson bracket is given by
\be
\label{1.1.1}
\{f,g\}=\Pi^{ij}(x)\partial_i f\partial_j g\ ,
\ee
where $\partial_i f:={\partial f\over\partial x_i}$ and the real quantities
$\Pi^{ij}(x)$ constitute the components of a 
 contravariant antisymmetric (real) 2-tensor
\be
\label{1.1.2}
\Pi=\Pi^{ij}(x)\partial_i \otimes \partial_j\ .
\ee
The latter is said to be a Poisson tensor and satisfies the condition
\be
\label{1.1.3}
\Pi^{li}\partial_l \Pi^{jk}+\Pi^{lj}\partial_l \Pi^{ki}+
\Pi^{lk}\partial_l \Pi^{ij}=0\ .
\ee
Conversely,
an antisymmetric (real) tensor 
$\Pi$ that fulfils (\ref{1.1.3})
defines a Poisson structure 
$\{\, ,\, \}$ by
\be
\label{1.1.4}
\{f,g\}=\Pi(df,dg)\ .
\ee
Straightforward  examples of Poisson manifolds are provided by symplectic manifolds,
e.g.~pairs $(\P,\omega)$ where $\omega$ is a  closed nondegenerate  2-form.
The Poisson tensor is in this case also  nondegenerate. It is given by the inverse
of the symplectic form, or explicitly by
\be
\label{1.1.5}
\Pi(i_X(\omega), i_Y(\omega))=\omega(X,Y)\ ,\ \forall X,Y\in\cal X(\P)\ , 
\ee
where $\cal X(\P)$ denotes the space of vector fields on $\P$. 
In fact, the property  
$d\omega=0$ guarantees that $\Pi$ (\ref{1.1.5}) satisfies the condition
(\ref{1.1.3}).

\medskip
Poisson manifolds $(\P,\pp)$ are precisely the mathematical structures
taken as models for finite dimensional classical systems, both in classical mechanics
and in classical statistical mechanics. The physical interpretation of the formalism
stands on the notions of {\em physical state} and {\em physical observable} and on
the relations between them. In the definition of states and observables below
we follow from the start an approach adapted to the \cst-algebras formalism.
\bd
\label{def1.1.2}
A physical state on a Poisson manifold $(\P,\pp)$ is a 
regular  Borel  probability measure on $\P$. The atomic measures, 
identified with points in $\P$, are said to be pure states, 
whereas the remaining ones are called mixed states.
\ed
Concerning observables, one typically considers the set of all real 
 $\cinf$ functions on $\P$. In this respect let us note the following.
\begin{itemize}
\item[] There is no inconvenient in considering complex functions, given that 
real functions are recovered as the invariant subset under involution.
Likewise, when considering the problem of quantization we will work
with complex algebras, taking into account the so-called {\em reality conditions}.
So, we will require that real functions belonging to the classical algebra are mapped
 under quantization to self-adjoint elements of the quantum algebra,  which is the same 
as requiring that the process of quantization maps the involution of functions to 
the natural involution of operators.
\item[] Aiming at the introduction of the \cst-algebra formalism, it is convenient
to work with a subspace
of $\cinfp$ which is also contained
on the \cst-algebra $C_0(\P)$  of continuous functions  vanishing  
at infinity\footnote{For locally compact $X$, $C_0(X)$ is the subset of those  $f\in C(X)$ with the property that for any $\epsilon >0$, there is a compact set $K_{\epsilon}\subset X$ such that $|f(x)|<\epsilon$ if $x\not\in K_{\epsilon}$. The \cst-norm of $C_0(X)$ is the supremum norm.}.
If $\P$ is compact the behaviour at infinity is not a question and one can take the full  space $\cinfp$. In case $\P$ is only locally compact, several choices appear possible,
for instance  $\cinf_0(\P)$,
the set of $\cinf$ functions vanishing at infinity, or $\cinf_{\rm c}(\P)$,
the set of   $\cinf$ functions with compact support. 
\end{itemize} 
Without claiming to establish what should be meant by {\em phyical observable},
let us adopt the following definition.
\bd
\label{def1.1.3}
A complete algebra of regular observables on a  Poisson manifold
$(\P,\pp)$ is a subspace $\ap\subset\cinf_0(\P)$ such that:
\begin{itemize}
\item[{\rm (}i\/{\rm )}] $(\ap,\pp)$ is a Lie  subalgebra of 
$(\cinfp,\pp)$.
\item[{\rm (}ii\/{\rm )}] $\ap$ is a dense $\ast$-subalgebra 
(with respect to the supremum  norm)
of $C_0(\P)$.
\end{itemize}
\ed
As examples of such complete algebras one can mention
 $\cinf_0(\P)$ and $\cinf_{\rm c}(\P)$, which are well defined in any circunstance. 
Nonetheless, other choices may be more convenient, in a given particular situation.
In any case, we will consider chosen a
complete algebra of  regular observables   $\ap$ such that
$\cinf_{\rm c}(\P)\subseteq\ap\subseteq\cinf_0(\P)$.

Such algebras are complete in the following sense: condition
({\em ii}\/) in definition \ref{def1.1.3} guarantees that
$\ap$ separates points in $\P$, i.e.
given $x\not=y$ em $\P$, there exists $f\in\ap$ such that $f(x)\not=f(y)$. 
\begin{itemize}
\label{nota1}
\item[] Let us clarify immediately the folowing. Although there are enough functions 
in a complete algebra of regular observables to define local coordinates, by no means
such an algebra contains all observables of physical interest, if $\P$ is noncompact.
The obvious example is provided by the standard global coordinate functions in 
 $\P={\rm T}^{\ast}\R$. Turning to quantization, the strategy will be to start with a convenient algebra of regular observables, seeking afterwards to extend the quantization map to other observables of interest, if necessary.
\end{itemize}
The quantities with direct physical correspondence are the (real) values of the 
pairing 
\be
\label{1.1.6}
(\mu,f)\mapsto \int d\mu(x)f(x)\ ,
\ee
for real functions $f$ and states $\mu$. Therefore, it is usually said  that the description
of the system in terms of states is ``dual'' to the description in terms of
observables. In the context of definitions \ref{def1.1.2} 
and \ref{def1.1.3}, this duality has a precise sense: the physical states show up as 
a subset of the unit ball in the dual of   $\ap$
(let us remind that $\ap$ is dense in $C_0(\P)$, and therefore the dual of $\ap$ 
coincides with the dual of  $C_0(\P)$). To be precise, let us introduce the following:
\bd
\label{def1.1.4}
A linear functional  $\varphi$ on a  $\ast$-algebra $\A$ is said to be positive
if $\varphi(a^{\ast}a)\geq 0\ \forall a\in\A$ (we write
$\varphi\geq 0$ to denote that $\varphi$ is  positive). A positive
linear functional on a  \cst-algebra is called a state (of the algebra)
if $\|\varphi\|=1$.
\ed
Note that a positive linear functional on a
\cst-algebra is necessarily continuous (see e.g.~\cite{B-R}).
\bp
\label{prop1.1.1}
Let $X$ be a locally compact Hausdorff space. Then the set of states
of the algebra $C_0(X)$ can be identified with the set of regular 
Borel  probability measures on $X$.
\ep
The proof of this result follows from the Riez-Markov theorem for locally  compact
spaces
(see \cite{R-S}), which  identifies the dual of $C_0(X)$
with the set  of finite complex regular  Borel measures in $X$, 
by means of the bijective correspondence
\be
\label{1.1.7}
\varphi\mapsto \mu_{\varphi}\ :\ \varphi(f)=\int f d\mu_{\varphi}\ ,
\forall f\in C_0(X).
\ee
The remaining nontrivial part of the proof consists in a typical functional analysis
argument, showing that 
 $\|\vf\|= \mu_{\vf}(X)$, which we will not present here. 

The physical states of the system $(\P,\pp)$ can therefore be seen as the states 
of the algebra $C_0(\P)$. The pure physical states 
(atomic measures) admit also important
characterizations in terms of the 
algebra $C_0(\P)$.
\bp
\label{prop1.1.2}
Given a locally compact Hausdorff space  $X$, there is bijective correspondence
between the following sets:
\begin{itemize}
\item[{\rm (}i\/{\rm )}] The set of atomic measures on $X$.
\item[{\rm (}ii\/{\rm )}] The set of nonnull linear functionals $\vf$ on the algebra $C_0(X)$
such that $\vf(ab)=\vf(a)\vf(b)\ \forall a,b\in C_0(X)$.
\item[{\rm (}iii\/{\rm )}] The set of irreducible representations of the algebra
 $C_0(X)$. 

\end{itemize}
\ep
The correspondence between ({\em i}\/) and ({\em ii}\/) is well known and constitutes 
part of Gelfand's theorem. The correspondence with 
({\em iii}\/) is easy to check: given that the algebra $C_0(X)$ is commutative, its irreducible
representations have dimension 1, and are therefore 
  \cst-algebras morphisms, $\varphi:C_0(X)\to 
\C$, i.e., belong to the set defined by ({\em ii}\/). 
On the other hand, it is clear that each atomic measure gives rise to such
a dimension 1 representation.

One can further show that the above sets defined by ({\em i}\/), ({\em ii}\/) and ({\em iii}\/)
are  equivalent to the set of {\em pure states of the algebra} 
 $C_0(X)$, with the following definition.
\bd
\label{def1.1.5}
A state $\vf$ on a \cst-algebra is said to be pure if the  conditions
$\vf\geq\chi\geq 0$ can only be fulfiled with $\chi=t\vf,\ t\in[0,1]$.
\ed

\section{Weyl-Moyal quantization}
\label{WM}
\subsection{Quantization map}
Let us consider the phase space $\P=T^{\ast}\R$, with local coordinates $(q,p)$, and its canonical symplectic
structure, defined by the form
\be
\label{1.2.1}
\omega = dq\wedge dp\ .
\ee
The associated  Poisson tensor is:
\be
\label{1.2.2}
\Pi=\partial_q \otimes \partial_p - \partial_p \otimes \partial_q\ ,
\ee
and therefore
\be
\label{1.2.3}
\{f,g\}=\partial_q f \partial_p g- \partial_p f \partial_q g\ .
\ee
We choose as algebra of observables $\ap$ the space 
$\S(T^{\ast}\R)\cong\S(\R^2)$ of Schwartz functions on 
$T^{\ast}\R\cong\R^2$. In this context, the Weyl-Moyal (W-M) quantization
consists of a family of linear maps  $\Qh$, $\hbar\in \R^+$,
from the Schwartz space to operators in $L^2(\R)$. Explicitly:
\be
\label{1.2.4}
\S(T^{\ast}\R)\ni f(q,p)\mapsto \Qh(f)\ :
\ee
\be
\label{1.2.5}
\Bigl(\Qh(f)\psi\Bigr)(q)=\int {dp\over 2\pi\hbar}e^{{ip\over \hbar}
(q-q\prime)}
f\Bigl({q+q\prime\over 2},p\Bigr)\psi(q\prime)dq\prime,
\ \psi\in L^2(\R).
\ee
We start by showing that the maps $\Qh$ have image on the subset of 
Hilbert-Schmidt operators. The operators $\Qh(f)$, $f\in  
\S(T^{\ast}\R)$, are in fact integral operators, of kernel
\be
\label{1.2.6}
K^f_{\hbar}(q,q\prime)=\int {dp\over 2\pi\hbar}e^{{ip\over \hbar}
(q-q\prime)}
f\Bigl({q+q\prime\over 2},p\Bigr)\ ,
\ee
which is clearly well defined and belongs to $\S(\R^2)$.
We get
\ba
\label{1.2.7}
\int dq dq\prime |K^f_{\hbar}(q,q\prime)|^2 & = & 
\int dq dq\prime\int {dp\over 2\pi\hbar}e^{{ip\over \hbar}
(q\prime-q)}
\bar f\Bigl({q+q\prime\over 2},p\Bigr)\cdot \nonumber \\
&\cdot &\int {dp\prime\over 2\pi\hbar}e^{{ip\prime\over \hbar}
(q-q\prime)}
f\Bigl({q+q\prime\over 2},p\prime\Bigr)\ .
\ea
With the new variables
\begin{equation}
v:={q+q\prime\over 2}\, , \ \
w:={q-q\prime\over \hbar}
\end{equation}
it follows that
\ba
\label{1.2.8}
\int dq dq\prime |K^f_{\hbar}(q,q\prime)|^2 & = & 
\int {dp\over 2\pi\hbar} dp\prime dv {dw\over 2\pi}
e^{iw(p\prime-p)}
\bar f(v,p)f(v,p\prime) \nonumber \\
&=&{1\over 2\pi\hbar}\int dvdp|f(v,p)|^2<\infty\ ,
\ea
for every $f\in\S(T^{\ast}\R)$, which shows that $\Qh(f)$
is an Hilbert-Schmidt operator. 

The smallest  \cst-subalgebra of  $B(L^2(\R))$
that contains the imagem of $\Qh$ coincides with the closure in the   uniform topology
of the set of Hilbert-Schmidt operators.  This is the space
${\cal K}(L^2(\R))$ of compact operators. Let us adopt ${\cal K}(L^2(\R))$
as common range of the maps $\Qh$.

It is straightforward to check that the quantization satisfies the reality conditions
\be
\label{1.2.9}
\Qh(\bar f)=\Qh^+(f)\ ,\ \forall \hbar\in \R^+,\ 
\forall f\in\S(T^{\ast}\R)\ .
\ee
In fact, it is obvious that the kernel $K^f_{\hbar}$ 
(\ref{1.2.6}) satisfies
\be
\label{1.2.10}
K^{\bar f}_{\hbar}(q,q\prime)= {{\bar K}}^f_{\hbar}(q\prime,q)\ ,
\ee
which is equivalent to (\ref{1.2.9}).

One can also show (see \cite{Lan}) that the W-M quantization (\ref{1.2.5})
satisfies the following conditions:
\ba
\label{1.2.11}
&(i)&\
\lim_{\hbar \to 0}\|\Q _{\hbar }(\{f,g\})-[\Q _{\hbar }(f)
,\Q _{\hbar }(g)]_{\hbar}\|=0\ .\\
\label{1.2.12}
&(ii)&\
\lim_{\hbar \to 0}\|\Q _{\hbar }(fg)-\Q _{\hbar }(f)\circ 
\Q _{\hbar }(g)\|=0\ .\\
\label{1.2.13}
&(iii)&\ \hbox{The maps}\ \hbar\mapsto\|\Qh(f)\|\
\hbox{are continuous in}\ \R^+,\ \forall f\, .\\
\label{1.2.14}
&(iv)&\
\lim_{\hbar \to 0}\|\Q _{\hbar }(f)\|=\|f\|\, (=
\hbox{sup}|f|)\ .
\ea
Condition ({\em i}\/) is the form in which
Dirac's quantization condition is implemented in this formalism: 
the classical Lie structure is not exactly preserved at the quantum level, 
but it is violated only by operators that tend to zero with $\hbar$.
Condition ({\em ii}\/) plays the same role with respect to the multiplicative
structure, ensuring that the algebraic relations between observables 
are recovered in the limit $\hbar \to 0$. 
Conditions ({\em iii}\/) and ({\em iv}\/) establish the continuity of the process
and provide, together with
 ({\em ii}\/), some control over the spectrum of the quantum operators.
 In particular, 
conditions ({\em iii}\/) and ({\em iv}\/) establish precisely the continuity
(near $\hbar=0$) of the spectral radius of the quantum operators.

Let us now clarify the relation between the W-M quantization and the usual
canonical quantization of the so-called Heisenberg algebra, i.e.~the Lie algebra 
generated by the coordinate functions $q$ and $p$. In the Dirac quantization, the observables
$q$ and  $p$ are mapped to operators $\hat q$ and  $\hat p$ in $L^2(\R)$, such that
\ba
\label{1.2.29}
\big({\hat q}\psi\bigl)(q)&=&q\psi(q)\\
\big({\hat p}\psi\bigl)(q)&=&-i\hbar{d\psi\over dq}(q).
\ea  
These operators are unbounded, and therefore cannot be defined for all $\psi\in L^2(\R)$.
It is standard procedure to restrict attention to the Schwartz subspace $\S(\R)\subset L^2(\R)$,
which is dense, belongs to the domain of both $\hat q$ and $\hat p$ and furthermore
remains invariant under the action of both operators.
If the action 
of $\Qh(f)$ (\ref{1.2.5}) is restricted to vectors  $\psi\in\S(\R)$, 
we see immediately that  $\Qh(f)$ remains well defined for a much larger 
class of functions
in  $T^{\ast}\R$ (see \cite{Fo} for a detailed discussion). 
In particular, $\Qh(q)$ and $\Qh(p)$
are well defined in $\S(\R)$ and coincide with the operators $\hat q$
and $\hat p$ above. Thus, the W-M quantization  is an extension of the canonical
quantization $\hat q$ and $\hat p$ of coordinate functions, to a large
class of observables $f(q,p)$.

\subsection{Positivity in the Weyl-Moyal quantization }
\label{pos}
In this section we address  the question of positivity in the 
Weyl-Moyal quantization. 

Let us recall that a self-adjoint operator $A$ is said to  be positive
if its expectation values are nonnegative, i.e.\ if $\langle\psi,A\psi\rangle\geq 0$, $\forall \psi$.
An equivalent condition is that   the spectrum of $A$
is a subset of $\R_0^+$.

Given the physical interpretation 
of observables, it is clearly desirable that, under a given quantization,  positive classical
observables (i.e.\ those that take only nonnegative values) 
are mapped to operators which are themselves positive. 
We show next that such condition is not fully satisfied in the W-M 
quantization.\footnote{See \cite{Fo} and \cite{Lan} for a general discussion.} 
However, Heisenberg's uncertainty relation helps in clarifying the situation,
showing why a weaker form of positivity is physically acceptable.  

For definitiness, let us consider the positive observables given
by gaussian functions in phase space, whose quantization is particularly simple.
Let then $f^{x_0}_{\alpha,\beta}$ denote  the  gaussian function:
\be
\label{1.2.15}
f^{x_0}_{\alpha,\beta}(q,p)=2\, e^{-{1\over 2}{(q-q_0)^2\over \a}}
 e^{-{1\over 2}{(p-p_0)^2\over \b}},
\ee
with arbitrary $x_0=(q_0,p_0)$ and $\alpha >0$, $\beta >0$.
The kernel  of the associated operador $\Qh\bigl(
f^{x_0}_{\alpha,\beta}\bigr)$ (\ref{1.2.5}) is easily found to be:
\be
\label{1.2.16}
K_{\hbar}^{f^{x_0}_{\alpha,\beta}}(q,q\prime)=\bar\chi(q)\chi(q\prime)
e^{-{1\over 8\a}({4\a\b \over \hbar^2}-1)(q-q\prime)^2},
\ee
where $\chi$ is an element of $\S(\R)$ given by
\be
\label{1.2.17}
\chi(q)=\left({2\b\over\pi\hbar^2}\right)^{1/4}
e^{-{1\over 4\a}(q-q_0)^2}
e^{-{ip_0\over \hbar}(q-q_0)} .
\ee
To address the question of positivity of
$\Qh\bigl(f^{x_0}_{\alpha,\beta}\bigr)$ let us then consider
the expectation values $\langle\psi,
\Qh\bigl(f^{x_0}_{\alpha,\beta}\bigr)\psi\rangle$, $\psi\in L^2(\R)$.
We get
\ba
\label{1.2.18}
\langle\psi,\Qh\bigl(f^{x_0}_{\alpha,\beta}\bigr)\psi\rangle&=&\\
&=&\int dqdq\prime\, \bar\chi(q)\bar\psi(q)
e^{-{1\over 8\a}({4\a\b \over \hbar^2}-1)(q-q\prime)^2}
\chi(q\prime)\psi(q\prime)\ .\nonumber
\ea
Let us prove that
$\langle\psi,\Qh\bigl(f^{x_0}_{\alpha,\beta}\bigr)\psi\rangle\geq 0$
 $\forall\psi$ if and only if $\a\b\geq\bigl(\hbar/2\bigl)^2$.
The conclusion is obvious for $\a\b =\bigl(\hbar/2\bigl)^2$.
For $\a\b >\bigl(\hbar/2\bigl)^2$ the conclusion follows from
the fact that, in this case, the gaussian function
in the integrand
can be written as the Fourier transform of a gaussian measure.
It remains to show that positivity fails for $\a\b <\bigl(\hbar/2\bigl)^2$,
i.e.\ that one can in this case find   $\psi\in L^2(\R)$ such that
$\langle\psi,\Qh\bigl(f^{x_0}_{\alpha,\beta}\bigr)\psi\rangle<0$.
To prove it, let us consider the family of vectors
\be
\label{1.2.19}
\psi_{\s}(q)=(q-q_0) 
e^{-{1\over 2\s}(q-q_0)^2}
e^{{ip_0\over \hbar}q},
\ee
with $\s>0$. From (\ref{1.2.17}) and (\ref{1.2.18}) we obtain
\ba
\label{1.2.21}
\langle\psi_{\s}
,
\Qh\bigl(f^{x_0}_{\alpha,\beta}\bigr)\psi_{\s}
\rangle
=
\\
\label{1.2.20}
%
=
&\left({2\b\over\pi\hbar^2}\right)^{1/2}
\int dq_1dq_2\, q_1 q_2 \, e^{-{1\over 2}({1\over\s} + 
{1\over 2\a})(q_1^2+q_2^2)}\cdot
e^{-{1\over 2}\Theta(q_1^2+q_2^2-2q_1 q_2)},\nonumber
\ea
where
\be
\label{1.2.22}
q_1:=q-q_0,\ \ q_2:=q\prime-q_0,\ \  \Theta:={1\over 4\a}\left({4\a\b \over \hbar^2}-1\right)\, .
\ee
Let us choose $\s$ such that
\be
\label{1.2.23}
{1\over\s}+{1\over 2\a}+2\,\Theta\not=0\, .
\ee
One can then write (\ref{1.2.21})
as a gaussian integral in $\R^2$:
\be
\label{1.2.24}
\langle\psi_{\s},\Qh\bigl(f^{x_0}_{\alpha,\beta}\bigr)\psi_{\s}\rangle=
\left({2\b\over\pi\hbar^2}\right)^{1/4}{2\pi\over D}
\int {dq_1dq_2\over(2\pi/D)}\, q_1 q_2 \, e^{-{1\over 2}
(q_1\, q_2)C^{-1}\bigl({}^{q_1}_{q_2}\bigr)}\ ,
\ee
where $C$ is the $2\times  2$ matrix such that
\be
\label{1.2.25}
C^{-1}:=\left(
\begin{array}{cc}
{1\over\s}+{1\over 2\a}+\Theta & - \Theta \\
- \Theta & {1\over\s}+{1\over 2\a}+\Theta
\end{array}
\right)
\ee
and 
\be
\label{1.2.26}
D:=\det C^{-1}=\big({1\over\s}+{1\over 2\a}\big)
\big({1\over\s}+{1\over 2\a}+2\,\Theta\big)\ .
\ee
The gaussian integral  (\ref{1.2.24}) is now trivial:
\be
\label{1.2.27}
\int {dq_1dq_2\over(2\pi/D)}\, q_1 q_2 \, e^{-{1\over 2}
(q_1\, q_2)C^{-1}\bigl({}^{q_1}_{q_2}\bigr)}=C_{12}={\Theta\over D}\, .
\ee
Putting it all together we finally get
\be
\label{1.2.28}
\langle\psi_{\s},\Qh\bigl(f^{x_0}_{\alpha,\beta}\bigr)\psi_{\s}\rangle=
\left({2\b\over\pi\hbar^2}\right)^{1/4}{2\pi\over D^2}\,\Theta\ .
\ee
Since one can obviously choose $\s >0$ compatible with  
(\ref{1.2.23}) and $\Theta<0$, it is clear 
that the operator $\Qh\bigl(f^{x_0}_{\alpha,\beta}\bigr)$,
associated with the positive observable $f^{x_0}_{\alpha,\beta}$,
is not positive for ${4\a\b\over\hbar^2}<1$.

It is interesting to analyse this  lack of positivity in the W-M quantization 
in light of Heisenberg's uncertainty relations. Note first that the observable  
$f_{\alpha,\beta }^{x_0}$ (\ref{1.2.15}) with
$\alpha \beta = (\hbar/2)^2$ is mapped precisely to the projector 
 (\ref{1.2.16}) onto the quantum  state $\chi $ (\ref{1.2.17}).
 This alows the semiclassical interpretation of
$f_{\alpha ,\beta }^{x_0}$ with
$\alpha \beta = (\hbar/2)^2$ as the  ``characteristic function
of the quantum state centered at $x_0$''. 
Less peaked  gaussian functions, i.e.\ with  $\alpha \beta > (\hbar/2)^2$
and therefore with a slow variation with respect to the quantum scale
 $\hbar/2$
are ``well quantized'', i.e.\ they are mapped to positive operators.
When it comes down to gaussian functions that probe regions  of the phase space 
of area less then $\hbar/2$ (which is the lower limit of the uncertainty relations), 
positivity is lost. Note however that, for $\alpha \beta < (\hbar/2)^2$, there
is no physical reason to require correspondence between (in particular the spectrum of) the
operator $\Qh\bigl(f^{x_0}_{\alpha,\beta}\bigr)$ and the observable $f^{x_0}_{\alpha,\beta}$.
In fact, precisely because those functions probe deep inside intrinsically
quantum domains in phase space, they are inaccessible to the classical
observer. The operators $\Qh\bigl(f^{x_0}_{\alpha,\beta}\bigr)$ in question,
if they are true physical observables, which is questionable, certainly have no classical limit.
 
\section{Strict quantization}
\label{strict}
Given a physical system admiting a classical mechanics description, by 
{\em quantization} one means finding a  ``corresponding''
quantum description. By hypothesis, the system in question
exhibits, under certain physical conditions determined by the
values of the physical observables involved,   a classical mechanical behaviour.
The 
correspondence between the classical model and the quantum
model is established at this limit:  the predictions of the classical model
should be a good approximation to the predictions of the 
 ``true quantum theory'' at the classical regime, i.e.\ when the system
 evolves subjected  to classical physical conditions. 

Although reasonably clear from the conceptual point of view,
the establishment of the classical limit of a quantum theory is  also a complex problem,
given the substantial differences between the formalisms of the two models,
classical and quantum.

In this sense, the emphasis on algebraic aspects constitutes a step
towards the formal approximation of the two models, useful both 
in the question of the classical limit and in the inverse problem, that of quantization.

Let us then consider a physical system, with which we associate
a Poisson manifold
 $\P $ and an  Hilbert space $\cal H$. As discussed in section \ref{class},
we assume as chosen a complete algebra of regular classical observables $\ap $.
Given that the quantum and the classical model describe the same system,
there should be a correspondence between functions
$f \in \ap $ and quantum operators, let's say  $\Q (f)$,
having $f$ a classical limit.
One expects the operators $\Q (f)$, $f \in \ap $, to be  bounded, and therefore
 $\Q $ should be a map between
$\ap $ and $B(\cal H)$, required to be linear and real, i.e., 
$\Q (\bar f)=\Q (f)^+$. 
The algebra of quantum observables is thus assumed to be
$B(\cal H)$, which is obviously complete, in the sense
that it acts irreducibly on $\cal H$. 

In this context, the observer deals with two algebras of observables:
the classical algebra, fitting  phenomena at the classical scale; and the quantum algebra, decribing, in principle, phenomena at any scale. A viable and useful perspective
consists in admitting the existence of a continuous family of algebras,
interpolating between the classical and the fully quantum domains.
This is in broad terms the quantization programme put forward by 
Rieffel \cite{Ri1} and Landsman \cite{Lan}, leading to 
the following definition.
\bd
\label{def1.3.1}
Let $\P $ be a Poisson manifold and $\ap $ a complete algebra of regular
observables on $\P $. Let
$\cal I\subset\R$ be a set containing zero as a limit point.
A strict quantization of $\ap $, labeled by $\cal I$, is a family of pairs 
$\{(A_{\hbar },\Qh )\}_{\hbar \in {\cal I}}$, where
each $A_{\hbar }$ is a  \cst-algebra and each $\Qh $ is a linear map
$\Qh :\ap \to A_{\hbar }$, with $A_0=C_0(\P )$, ${\cal Q}_0(f)=f,\ 
\forall f\in\ap$, such that:
\begin{itemize}

\item[{\rm (}i\/{\rm )}] $\Qh(\bar f)=\Qh^+(f)\ ,\ \forall \hbar\in I,\ 
\forall f\in\ap\, .$

\item[{\rm (}ii\/{\rm )}] $\lim_{\hbar \to 0}\|\Q _{\hbar }(\{f,g\})-
[\Q _{\hbar }(f),\Q _{\hbar }(g)]_{\hbar}\|=0\, .$

\item[{\rm (}iii\/{\rm )}] $\lim_{\hbar \to 0}\|\Q _{\hbar }(fg)-
\Q _{\hbar }(f)\circ \Q _{\hbar }(g)\|=0\, .$

\item[{\rm (}iv\/{\rm )}] $\lim_{\hbar \to 0}\|\Q _{\hbar }(f)\|=\|f\|\,  .$

\end{itemize} 

\ed
Conditions ${\rm (}ii\/{\rm )}$ to ${\rm (}iv\/{\rm )}$ establish 
the sense in which the classical limit is understood or, from the point of view of 
quantization, the conditions that the maps $\Qh $ should fulfil
in order to ensure correspondence with the classical theory.
Condition 
${\rm (}iii\/{\rm )}$ replaces the so-called 
 von Neumann condition on the  preservation of the multiplicative structure. 
Condition
${\rm (}ii\/{\rm )}$ is the implementation, in this formalism, of Dirac's
original ideia that the quantum correspondent of the  Poisson bracket is the
quantum Lie bracket  
$[\, ,\,]_{\hbar}$. Condition
${\rm (}iv\/{\rm )}$ gives some control over the spectral radius of the operators,
ensuring in particular that the quantum spectrum is
not   radically different from the classical spectrum.

The perfect example of a strict quantization is the Weyl-Moyal quantization.
In this particular case, the maps
$\Qh :\S (\R )\to {\cal K} (L^2(\R ))$ are bijective and it is therefore possible,
using the inverse maps $\Qh^{-1}$, to transpose the multiplicative
structures over to $\S (\R )$, i.e., to define a family of \cst-products, 
say $\star _\hbar $, on $\S (\R )$, thus obtaining a deformation of the commutative
algebra  $\ap $.
\bd
\label{def1.3.2}
A strict quantization $\{(A_{\hbar },\Qh )\}_{\hbar \in {\cal I}}$
is said to be a strict deformation quantization if $\Qh (A_0)$ is a
subalgebra and the  maps $\Qh $ are injective. 
\ed
%


\section{Smooth groupoids}
\label{gru}
Some interesting deformations of classical algebras, and in particular 
Landsman's quantization of the cotangent bundle of a Riemannian manifold,
are naturally associated with groupoid convolution algebras.
We review here very briefly the necessary notions, 
following  \cite{Co} and \cite{SW}.

A \gg\ $G$ can be seen as a generalization of the notion of group.
In a group every element can be combined with each other, i.e.~there is a map
$G \times G \to G$. In a \gg\ ones drops the hypothesis that the map is defined
for every pair of elements; it is only assumed the existence of a binary operation
on a subset, say $G^{(2)}$, of $G \times G$. 
\bd
\label{def2.1.1}
A \gg\ is a (concrete) category $G$ such that all the arrows in the category have an inverse.
The elements of the \gg\ are the arrows of the category,  the composition of which 
defines the binary operation on the \gg. 
\ed
We present next some examples of groupoids.
In what follows, we identify the set Obj$G$ of objects of the category
with the set of identity arrows  and denote  by $G$ both the category 
and the set {\rm Morf}$G$ of its morphisms, or arrows.
We use still the following notation:
${\rm Hom}\,[x,y]$ denotes the set of morphisms from $x$ to $y$;
$s$ (resp.~$r$) denotes the map that applies $g\in {\rm Hom}\,[x,y]$ into $x$ (resp.~$y$).

\medskip
{\parindent=0pt
{\bf Example 1}. A group is a category with the identity as the only  object.
The elements of the group are the arrows of the   category, composition being
the group operation. Since all arrows are invertible, a group is a \gg.}

\medskip
{\parindent=0pt
{\bf Example 2}. A vector bundle ($E,V,M,\Pi $) with fiber
$V$ over a manifold $M$ is a \cc\ whose objects are the points of $M$. 
The arrows of the \cc\ are the elements of $V$ at each point of
 $M$, i.e., ${\rm Hom}\,[x,y]=\emptyset$ if $x \neq y$ and
${\rm Hom}\,[x,x]=\Pi _xE \cong V$. Composition of arrows is defined
by vector sum in the fiber, i.e., $(x,X)(x,Y)=(x,X+Y)$, $x \in M$,
$X,Y \in V$.}

\medskip
{\parindent=0pt
{\bf Example 3}. Given a set $X$, the product 
$X\times X$ is a groupoid with the following category structure: the objects
of the  category are the points of $X$, the arrows are the elements of
$X\times X$, i.e.~${\rm Hom}\,[y,x]=\{(x,y)\}$. The composition of arrows is given by 
 $(x,y)(y,z)=(x,z)$.}

\medskip
{\parindent=0pt
{\bf Example 4}. Given a set $X$, a group $\Gamma$ and a right action
$\alpha:X\times \Gamma\to X$, one can obtain the  semidirect product \gg\
{\bf } $G=X>\!{\lhd}_{\alpha}\Gamma$ as follows. The set of objects coincides with $X$. 
The arrows constitute the set $X\times\Gamma$,
with $(x,\gamma)\in{\rm Hom}\,[{\alpha}_{\gamma}(x),x]$. Combination of arrow is given by:
$(x,\gamma_1)(y,\gamma_2)=(x,\gamma_1\gamma_2),
\ {\rm if}\ {\alpha}_{\gamma_1}(x)=y$. The inverse of the arrow
$(x,\gamma)$ is $({\alpha}_{\gamma}(x),\gamma^{-1})$.}

\medskip
Let us now consider the introduction of a compatible differential structure
on a groupoid \cite{Co}.
\bd
\label{def2.1.2}
A smooth \gg\  is a \gg\ $G$ such that:
\begin{itemize}
\item[{\rm (}i\/{\rm )}] $G$, {\rm Obj}G and the set
$G^{(2)}\subset G\times G$ of pairs of combinable arrows are smooth manifolds.
\item[{\rm (}ii\/{\rm )}] The inclusion {\rm Obj}$G\to G$, 
the composition of arrows 
$G^{(2)}\to G$ and the inversion of arrows are smooth maps.
\item[{\rm (}iii\/{\rm )}] The maps\ $\, r,s:G\to\, ${\rm Obj}G 
are submersions.
\end{itemize}
\ed
One can now construct convolution algebras 
and finally a \cst-algebra associated with
a smooth \gg\ and a family of measures, as follows.

Let then $G$ be a smooth \gg.  
For each $x\in{\rm Obj}G$, consider the sets $G^x:={\cup}_{y\in {\rm Obj}G}{\rm Hom}\, [y,x]$ 
and
$G_x:={\cup}_{y\in {\rm Obj}G}{\rm Hom}\, [x,y]$,
called {\it r\/}-fibre and {\it s\/}-fibre, respectively. 
These fibres inherited a locally compact topology, induced from
the manifold structure of $G$ \cite{Co}. 

For each $g\in G$  there is a map
$\Theta_g:G^{s(g)}\to G^{r(g)}$, defined by $\Theta_g(g')=gg'$, which establishes
a bijection between $G^{s(g)}$ and $G^{r(g)}$ (since every arrow $g$ is invertible). 

A family of measures $\mu^x$ on the  {\it r\/}-fibres $G^x$ is said to be a
{\em  Haar system} if it satisfies the compatibility conditions
$\mu^{r(g)}=\bigl(\Theta_g\bigr)_{\ast}\mu^{s(g)}$, $\forall g\in G$,
where $\bigl(\Theta_g\bigr)_{\ast}\mu^{s(g)}$ is the {push-forward}
of the  measure $\mu^{s(g)}$, with respect to the map  $\Theta_g$.
Finally, note that the map $g\mapsto g^{-1}$ establishes also a bijection between 
$G^x$ and $G_x$, for every
$x\in{\rm Obj}G$.  By means of these bijections, a Haar system defines also
a family of measures on the  {\it s\/}-fibres $G_x$. 
\bd
\label{def2.1.3} Let G be a smooth \gg\ equipped with a Haar system of measures.
The following defines a convolution on
the space of $C^{\infty}$ functions on G with compact support:
\be
\label{2.1.1}
(F_1\star F_2)(g)=\int_{G^{r(g)}}\, F_1(h)F_2(h^{-1}g)d\mu^{r(g)}(h),\ \
 F_1,F_2\in C^{\infty}_c(G).
\ee
On the same space, an involution is defined by
\be
\label{2.1.1a}
F(g)\mapsto{\bar F}(g^{-1})\ ,\ \ F\in C^{\infty}_c(G).
\ee
\ed
The convolution algebra $C^{\infty}_c(G)$ admits natural $\ast$-representations,
one per each  $x\in{\rm Obj}G$. In fact, let us consider the  Hilbert spaces
$L^2(G_x,\mu _x)$, $x\in{\rm Obj}G$. The (involutive) representations
 $\Bigl(L^2(G_x,\mu _x),\pi _x\Bigr)$  are defined by
\be
\label{2.1.2} 
\Bigl(\pi _x(F)\psi \Bigr)(g)=\int_{G^{r(g)}}\, 
F(h)\psi (h^{-1}g)d\mu^{r(g)}(h)\ ,
\ee
where $F\in C^{\infty}_c(G)$, $\psi \in L^2(G_x,\mu _x)$ and $g \in G_x$.
\bd
\label{def2.1.4} Let G be a smooth \gg\ with Haar system $\{\mu ^x\}_{x\in{\rm Obj}G}$.
The associated (reduced) \cst-algebra  $C^{\ast}_r(G)$ is the completion
of the convolution algebra  $C^{\infty}_c(G)$ with respect to the norm
 $\|F\|:={\rm sup}_{x\in{\rm Obj}G}\|\pi _x(F)\|$.
\ed
Let us analyse again  the previous examples, each of which \gg\ is now equipped
with a natural differential structure.

\medskip
{\parindent=0pt
{\bf Example 1a}. In a  group there is only one object, and therefore we have only
one {\it r\/}-fibre and one {\it s\/}-fibre, 
both coincident with the group itself.
In this case a locally compact topology is sufficient to construct
the \cst-algebra, which coincides with the convolution algebra on the group
for a given Haar measure. Consider in particular the additive group $\R $ with the
 Lebesgue measure: the obtained \cst-algebra is the 
 Fourier transform of the multiplicative algebra  $C_0(\R )$.}

\medskip
{\parindent=0pt
{\bf Example 2a}. We consider the particular case of a tangent bundle  $TM$ 
of a $n$-dimensional Riemannian manifold $(M,\tt g)$. 
Let us fix a local coordinate system  $(q^1,\dots,q^n)$ in $M$, colectively denoted by
$q$. At each point $q\in M$ the  {\it r\/}-fibre
$G^q$ coincides with  $T_qM$, which in turn can be identified with 
$\R^n$ by $X=(X^1,\ldots,X^n)\mapsto\sum X^i\partial_{q^i}$.
On the {\it r\/}-fibres  we consider the measure
$d\mu ^q(X)=\sqrt {{{\rm det}\tt g}(q)}d^nX$, 
where $d^nX$ is the Lebesgue measure and  {\tt g} is the metric.
Convolution is given by    integration on the fibre at each point of $M$:
\be
\label{2.1.3} 
(F_1\star F_2)(q,X)=\int_{T_qM}\, F_1(q,Y)F_2(q,X-Y)d\mu ^q(Y)\, .
\ee
The obtained \cst-algebra is   isomorphic to the  multiplicative algebra
$C_0(T^*M)$, by  Fourier transform   $\cal F$ on the fibre:
\be
\label{2.1.4} 
({\cal F}F)(q,\xi ):=\int_{T_qM}\, {\rm e}^{-i\xi X}F(q,X)d\mu ^q(X),
\ee 
where $(q,\xi )\in T^*M$, ${\cal F}F\in C_0(T^*M)$.}

\medskip
{\parindent=0pt
{\bf Example 3a}. Let $(M,\tt g)$ be a $n$-dimensional Riemannian manifold and consider the
\gg\ $M\times M$.
Every {\it r\/}-fibre and {\it s\/}-fibre is isomorphic to
$M$. On each fibre, we consider the measure  $d\mu (q)=\sqrt {{\rm det}{\tt g}(q)}d^nq$,
for some local coordinate system  on $M$. The convolution is
\ba
(F_1\star F_2)(q,q')&=&\int_{G^q}\, F_1(q,q'')F_2\bigl((q,q'')^{-1}(q,q')
\bigr)d\mu (q'')\\
 &=&\int_M\, F_1(q,q'')F_2(q'',q')d\mu (q''),
\ea
where one can recognize immediately the convolution of kernels of integral
operators in $L^2(M,\mu )$. In fact, the algebra $C^{\ast}_r(M\times M)$
is isomorphic to the algebra ${\cal K}\bigl(L^2(M,\mu )\bigr)$ of compact operators
in $L^2(M,\mu )$ \cite{Va}.}

\medskip
{\parindent=0pt
{\bf Example 4a}. In this case we analyse an example directly related 
to the Weyl-Moyal quantization.
Consider a family of actions $\alpha ^{\epsilon }$ of 
$\R$ on $\R$, labeled by a real number $\epsilon$.
For each  $\epsilon$ the actions are 
$\alpha_y^{\epsilon}(x)=x+\epsilon y$. Independently of
$\epsilon $, the {\it r\/}-fibres and {\it s\/}-fibres of the semidirect product 
$\R >\!{\lhd}_{\alpha ^{\epsilon }}\R$ can be identified with $\R$. Let us then introduce the  Lebesgue measure on each fiber. The convolution is then given by
\ba
(f\star g)(x,y)&=&\int_{G^x}\, f(x,z)g\bigl((x,z)^{-1}(x,y)\bigr)dz\\
 &=&\int_{\R}\, f(x,z)g\bigl((x+\epsilon z,-z)(x,y)\bigr)dz\\
 &=&\int_{\R}\, f(x,z)g(x+\epsilon z,y-z)dz.
\ea
Concerning the action of the elements  of the algebra on the
Hilbert spaces $L^2(G_u)$, let us distinguish the cases  $\epsilon =0$ and
$\epsilon \neq 0$. For $\epsilon >0$ the action does not depend on the
{\it s\/}-fibre. It is defined on $L^2(\R)$ by
\be
\label{2.1.5} 
\bigl(\pi (f)\psi \bigr)(x)=\int_{\R}\, f(x,y)\psi (x+\epsilon y)dy,\ \psi \in L^2(\R).
\ee 
For $\epsilon =0$ we get the representations
$\bigl\{\bigl(\pi _u,L^2(\R)\bigr)\bigr\}_{u\in \R}$:
\be
\label{2.1.6} 
\bigl((\pi _uf)\psi \bigr)(x)=\int_{\R}\, f(u,y)\psi (x-y)dy.
\ee 
We recognize for $\epsilon =0$ the Fourier transform (in the second variable)
of the  multiplicative algebra
$C_0(\R \times \R)$. In the case $\epsilon >0$ the nontrivial action
 of $\R$ on $\R$ deforms the convolution
algebra  in a way that
corresponds precisely to the Weyl-Moyal deformation,  see  section \ref{qt} below.}

\section{Tangent groupoid}
\label{tg}
This section is dedicated to Landsman's quantization of the cotangent bundle of 
a Riemannian manifold $Q$ \cite{Lan1}.
Although this construction can be described without reference to Connes' 
tangent groupoid \cite{Co}, we adhere to the tangent groupoid perspective right from the start,
since it is quite natural, both from the geometric and the algebraic viewpoints.
We start by showing how the simplest case, that of $Q=\R$, fits in this framework.
\subsection{Weyl-Moyal quantization revisited}

The Weyl-Moyal  quantization admits a reformulation in terms
of  the so-called tangent groupoid.
As we will see shortly, the classical algebra and the quantum  algebras
appear unified, as elements of the same algebra of functions on the tangent groupoid.

First note that the W-M quantization maps $\Qh$ (\ref{1.2.5}) can be naturally
split into  three distinct maps. Consider the
 Fourier transform
\be
\label{1.4.1}
{\cal F}:\S(T^{\ast}\R)\rightarrow\S(T\R)
\ee
\be
\label{1.4.2}
f(q,p)\mapsto \tilde f(q,v)=\int {dp\over 2\pi}\, e^{ip v}
f\bigl(q,p\bigr)
\ee
and the representation $\pi:\S(\R\times\R)\rightarrow {\cal K}(L^2(\R))$ 
of $\S(\R\times\R)$ functions as kernels of integral  operators.
It is then clear that the maps 
$\Qh:\S(T^{\ast}\R)\rightarrow {\cal K}(L^2(\R))$ 
(\ref{1.2.5}) are obtained as the composition
\be
\label{1.4.3}
\Qh=\pi\circ\varphi_{\hbar}\circ{\cal F}\ ,
\ee
where the map $\varphi_{\hbar}:\S(T\R)\rightarrow \S(\R\times\R)$
is defined by
\be
\label{1.4.4}
\bigl( \varphi_{\hbar}\tilde f\bigr)(x,y)={1\over \hbar}\tilde f
\bigl({x+y\over 2},{x-y\over \hbar}\bigr).
\ee
The importance of this  decomposition is that it isolates the nontrivial
map $\varphi_{\hbar}$, in which is effectively present the 
deformation of the  algebraic structure. In fact, the Fourier transform
is a  morphism  from the multiplicative algebra $\S(T^{\ast}\R)$ to the
 convolution algebra (with respect to the second variable) in $\S(T\R)$.
 This is, in turn, the algebra naturally associated with  the \gg\ structure
 of $T\R$.
As we have seen above, $\R\times\R$ is also a \gg, 
and the map $\pi$ is precisely a morphism from the associated \gg\ algebra to the
algebra ${\cal K}(L^2(\R))$ of compact operators.

The  W-M maps are therefore  naturally decomposed into a couple of morphisms
and a map between \gg\ algebras, the 
deformation $\varphi_{\hbar}$:
$$
\begin{array}{ccc}
 \S(T^{\ast}\R) & \stackrel{\Qh}{\rightarrow} & {\cal K}(L^2(\R)) \\
 {\cal F}\downarrow & {} & \uparrow\pi \\
 C_r^{\ast}(T\R) & \stackrel{\varphi_{\hbar}}{\rightarrow} & 
 C_r^{\ast}((\R\times\R)\times\{\hbar\})\ ,
\end{array}
$$
where $C_r^{\ast}(T\R)$ denotes the $C^{\ast}$-algebra of the \gg\  $T\R$
and $C_r^{\ast}((\R\times\R)\times\{\hbar\})$ denotes the $C^{\ast}$-algebra of the \gg\
 $(\R\times\R)\times\{\hbar\}\cong \R\times\R$.

The crucial point is that the transformations  $\varphi_{\hbar}$ 
are induced by an identification of
 $T\R$ with $\R\times\R$. Let us consider the map
$\phi:T\R\rightarrow\R\times\R$ defined by
\be
\label{1.4.5}
T\R\ni(q,v)\stackrel{\phi}{\mapsto}
(q+{1\over 2}v,q-{1\over 2}v)\, .
\ee
Consider still the family of maps 
$\phi_{\hbar}:T\R\rightarrow(\R\times\R)\times\{\hbar\}$ obtained from the previous one:
\be
\label{1.4.5a}
\phi_{\hbar}(q,v)=\phi(q,\hbar v)= (q+{\hbar\over 2}v,q-{\hbar\over 2}v)\, .
\ee
It is then clear  that
\be
\label{1.4.6}
\bigl(\varphi_{\hbar}\ft\bigr)(x,y)={1\over\hbar}
\ft\bigl(\f_{\hbar}^{-1}(x,y)\bigr)\, .
\ee
The maps $\phi_{\hbar}$ (\ref{1.4.5a}) allow the construction of a
manifold with boundary, the so-called tangent groupoid \cite{Co}, as follows.
Consider  first the product manifold $G_{\R}^1:=(\R\times\R)\times]0,1]$.
This is also a \gg: two elements $(x,y,\hbar)$ and
$(x',y',\hbar')$ can be combined if and only if they belong to the same leaf
 $(\R\times\R)\times\{\hbar\}$, i.e. if $\hbar=\hbar'$.
The associated \cst-algebra  turns out to be $C_r^{\ast}(G_{\R}^1)\cong
C_0(]0,1])
\otimes C_r^{\ast}(\R\times\R)\cong
C_0(]0,1])
\otimes {\cal K}(L^2(\R))$.
It is  interesting to note that $G_{\R}^1$ can be seen as the space 
of secant lines  to $\R$,  or more precisely of finite difference operators. 
In fact, the elements $(x,y,\hbar)\in G_{\R}^1$ define
elements of the  dual of the space $C^1(\R)$ of differentiable functions in $\R$,  by
$$(x,y,\hbar)\mapsto {f(x)-f(y)\over\hbar}\ ,\ f\in C^1(\R)\ .$$
The closure of this open set in ${C^1(\R)}^{\ast}$ is the  
tangent groupoid \cite{Co,VaC}. 

An   explicit construction of the 
 tangent groupoid, both as a manifold and as a groupoid, is the folowing \cite{Co}
(see also \cite{VaC,Va}). Let us consider the union 
$G_{\R}:=G_{\R}^1\cup G_{\R}^2$, where $G_{\R}^2:=T\R$. $G_{\R}$ is a \gg\ with the obvious 
structure of union of groupoids.
The structure of manifold with boundary is defined  by the maps
 $\phi_{\hbar}$ (\ref{1.4.5a}), which give coordinates in
 $G_{\R}$. In fact, as a manifold, the tangent \gg\ $G_{\R}$ is
diffeomorphic to $T\R\times[0,1]$.
The announced coordinate system 
in $G_{\R}$ is given by the transformation
\be
\label{1.4.7}
\F:T\R\times[0,1]
\rightarrow G_{\R}
\ee
such that
$$
\F(q,v,\hbar)=\left\{ \begin{array}{ll}
(q+{\hbar\over 2}v,q-{\hbar\over 2}v,\hbar) & \mbox{if\ $\ \hbar>0$} \\
(q,v) & \mbox{if\ $\ \hbar=0$.}\end{array}
\right.
$$
The transformation  $\F$ is a diffeomorphism when restricted to
$T\R\times]0,1]$
and maps the boundary of $T\R\times[0,1]$
to the boundary of $G_{\R}$.

The \cst-algebra  $C_r^{\ast}(G_{\R})$ associated with the tangent groupoid $G_{\R}$ 
is formed by pairs
$\bigl(\{k_{\hbar}\}_{\hbar\in]0,1]}
,\ft\bigr)$,
where $k_{\hbar}
\in C_0(\R\times\R)$, $\ft\in C_0(T\R)$, subject to the continuity 
condition at the boundary, i.e.
\be
\label{1.4.8}
\lim_{\hbar\to 0}k_{\hbar}\bigl(q+{\hbar\over 2}v,q-{\hbar\over 2}v\bigr) 
=\ft(q,v)\ .
\ee 
This continuity 
condition  is in fact a  quantization condition, imposing
that any  element of $C_r^{\ast}(G_{\R})$ is a family of quantum operators
having $\ft(q,v)$ (or its inverse Fourier transform $f(q,p)$) as a limit. 

The above continuity condition at the boundary still leaves a great deal of freedom
as to the choice of quantum operators to be associated with a given classical observable,
and therefore the quantization maps are not fixed.
However, the construction of the \gg\ itself suggests the construction
of well determined linear maps
  $\S(T\R)\rightarrow C_0(\R\times\R)$.
Let then $\ft(q,v)$ be an element of $\S(T\R)$. Consider the element
$\tilde F:=\{\ft_{\hbar}\}_{\hbar\in[0,1]}$
of
$C_r^{\ast}(T\R\times [0,1])$
given by $\ft_{\hbar}=\ft\, ,\ 
\forall\hbar$.
The element ${1\over \hbar}({\F^{-1}})^{\ast}\tilde F$ of
$C_r^{\ast}(G_{\R})$ provides then the required 
quantization. The composition of this map with the 
 Fourier transform and the representation $\pi$ finally gives the 
 Weyl-Moyal quantization.

\subsection{Cotangent bundle of a  Riemannian manifold}
Following Landsman \cite{Lan, Lan1}, Connes \cite{Co} and also reference \cite{VaC},
we present in this section  the strict quantization of the 
most common type of
phase space in physical applications, which is the 
 cotangent bundle
$T^{\ast}Q$ of some Riemannian manifold $Q$.  We start with the construction of the tangent \gg\
$G_Q$, which generalizes the construction of the previous section.
We introduce first the algebraic structure of the tangent \gg\ $G_Q$, followed by
its manifold structure.

Let then $(Q,\tt g)$ be a $n$-dimensional Riemannian manifold, with metric $\tt g$. The associated tangent \gg\ is a disjoint union of groupoids,
 $G_Q=((Q\times Q)\times ]0,1])\, \dot{\cup}\, TQ$,
formed by the \gg\ $TQ$ and by a family of copies of the \gg\
 $Q\times Q$. The tangent bundle $TQ$ is a smooth \gg, 
equipped with a Haar system of measures $d\mu^q$ on the  fibres $T_qQ$: 
\be
\label{2.2.1}
d\mu^q(v)=d^nv\sqrt{{\rm det}\, {\tt g}(q)}\ ,
\ee
where $(q,v)$ denotes local coordinates on $TQ$.
The    convolution algebra of $TQ$ is determined by the expression
\be
\label{2.2.2}
(f\star g)(q,v)=\int d\mu^q(v') f(q,v')g(q,v-v')\ .
\ee
The product $Q\times Q$ is also a smooth \gg, with measure
\be
\label{2.2.3}
d\mu(q)=d^nq\sqrt{{\rm det}\, {\tt g}(q)}
\ee
on the fiber $Q$. The product  $(Q\times Q)\times ]0,1]$ is again a smooth \gg, with
the  product manifold structure and the following \gg\  structure.
The elements $(x,y,\hbar)$ and $(x',y',\hbar')$ can be combined
if and only if $\hbar=\hbar'$ and $y=x'$, and in that case
$(x,y,\hbar)(y,y',\hbar)=(x,y',\hbar)$. The  {\it r\/}-fibres and  
{\it s\/}-fibres of $(Q\times Q)\times ]0,1]$ both coincide with $Q\times\{\hbar\}\cong Q$, and the convolution algebra is given by
\be
\label{2.2.4}
(f\star g)(x,y,\hbar)=\int d\mu(z) f(x,z,\hbar)g(x,y,\hbar)\ .
\ee
Finally, the disjoint union
$G_Q=((Q\times Q)\times ]0,1])\, \dot{\cup}\, TQ$
acquires a natural \gg\ structure, in the sense that elements of 
$(Q\times Q)\times ]0,1]$ (resp. $TQ$) combine only amongst themselves.

As in the previous section, which corresponds to $Q=\R$, $G_Q$ becomes a smooth \gg\ 
\cite{Co,VaC,Va}  when equipped with the topology of a manifold with boundary.
Generalizing the previous procedure, we present next  a map
from an open set $U\subset TQ\times ]0,1]$ to $G_Q$,
which defines the boundary of $G_Q$. 

Let us fix a local coordinate system $q$ on $Q$. This gives us also
coordinates
$(q,q')$ on $Q\times Q$ and a basis $(\partial_q,\partial_{q'})$
on each space $T_{(q,q')}(Q\times Q)$. Consider the diagonal embedding
$\Delta:Q\to  Q\times Q$ given by $\Delta(q)=(q,q)$. At each point
$\Delta(q)$ the metric $\tt g\oplus\tt g$ on $Q\times Q$ allows a decomposition
of $T_{(q,q)}(Q\times Q)$ vectors in tangent and normal parts, i.e.
$$T_{(q,q)}(Q\times Q)=\Delta_{\ast}T_qQ\oplus(\Delta_{\ast}T_qQ)^{\perp}\, ,$$
where $\Delta_{\ast}T_qQ$ is the push-forward of the tangent space and $(\Delta_{\ast}T_qQ)^{\perp}$ is its orthogonal complement.
The union  $\cup_{q\in Q}(\Delta_{\ast}T_qQ)^{\perp}$ is a subbundle of the restriction
of $T(Q\times Q)$ to $\Delta(q)$,
whose fibres are normal to 
$\Delta(q)$ at each point. This is the normal bundle associated with $\Delta$,
hereafter denoted  by $N^{\Delta}Q$.

Clearly,  the elements of $\Delta_{\ast}T_qQ$
are of the form $(X_q,X_q)$, with $X_q\in T_qQ$ and in the same way
the elements of $(\Delta_{\ast}T_qQ)^{\perp}$ can be written in the form
$(X_q,-X_q)$, with $X_q\in T_qQ$. One can therefore build a map
$\eta:TQ\to N^{\Delta}Q$, given by 
\be
\label{2.2.5}
\eta(q,X_q)=\Bigl(\Delta(q),{1\over 2}X_q,-{1\over 2}X_q\Bigr)\, .
\ee
The transformation $\eta$ can now be combined with the exponential
map defined by normal geodesics at $\Delta(q)$. Let then
$W_1$ be an open set in $N^{\Delta}Q$ 
where the exponencial map is defined and let $U_1\subset Q\times Q$
denote the image of $W_1$. Consider the transformations $\f:V_1\to U_1$, 
$\f=\exp\circ\eta$, where $V_1:=\eta^{-1}(W_1)\in TQ$. Explicitly
\be
\label{2.2.6}
TQ\ni (q,X_q)\stackrel{\eta}{\mapsto}\Bigl(\Delta(q),{1\over 2}X_q,
-{1\over 2}X_q\Bigr)
\stackrel{\exp}{\mapsto}\Bigl(\exp_q({1\over 2}X_q),
\exp_q(-{1\over 2}X_q)\Bigr)\ .
\ee
Let us define still the maps 
\ba
\label{2.2.7}
\f_{\hbar}:V_{\hbar}:={1\over \hbar}V_1\to & V_1 & \stackrel{\f}{\to}
U_{\hbar}\cong U_1\times \{\hbar\}\\
TQ\ni (q,X_q)\mapsto & (q,\hbar X_q) &  \mapsto \Bigl(\exp_q({1\over 2}
\hbar X_q),\exp_q(-{1\over 2}\hbar X_q),\hbar\Bigr).
\ea
Finally, consider the manifold $TQ\times[0,1]$ with its product structure
and  its boundary  $TQ\times\{0\}\cong TQ$. Let $U$ be the open set in
$TQ\times[0,1]$ defined  by $U:=\bigl(X_{\hbar\in]0,1]}V_{\hbar}\bigr)
\times\bigl(TQ\times\{0\}\bigr)$, which contains the boundary. 
The transformation
$\Phi:U\to G_Q$ defined by
$$
\F(q,X_q,\hbar)=\left\{ \begin{array}{ll}
\f_{\hbar}(q,X_q) & \mbox{if $\ \hbar>0$} \\
(q,X_q) & \mbox{if $\ \hbar=0$}\end{array}
\right.
$$
is a diffeomorphism when restricted to $TQ\times]0,1]$ and maps
$TQ$ to $TQ$, thus defining a coordinate system on an open set in $G_Q$ 
containing the boundary. This concludes the description of the tangent \gg\  $G_Q$.

Let us then describe the quantization of the symplectic  manifold $\P:=T^{\ast}Q$.
Consider first the measure
\be
\label{2.2.8}
d\mu_q(p)={d^np\over (2\pi)^n\sqrt{{\rm det}\, {\tt g}(q)}}
\ee
on the  fibres $T_q^{\ast}Q$ of $T^{\ast}Q$, where $(q,p)$ is a local coordinate system. The Fourier transform
on the fibre maps functions $f(q,p)$ on $T^{\ast}Q$ to functions
$\tilde f(q,v)$ on $TQ$:
\be
\label{2.2.9}
\tilde f(q,v):=\int d\mu_q(p)\, e^{ipv}f(q,p)\ .
\ee
Let us adopt as a complete algebra of regular observables $\ap$ the
 subalgebra 
of the functions
$f\in C_0^{\infty}(\P)$ such that $\tilde f\in C_{\rm c}^{\infty}(TQ)$.

The quantization maps are defined as follows. 
For a given observable $f\in \ap$, let $\hbar(f)$ be  a real number such that
${\rm supp}\tilde f\subset V_{\hbar(f)}$.
Then, for every $\hbar\leq\hbar(f)$, 
\be
\label{quantLan}
K^f_{\hbar}:={\hbar}^{-n}
\bigl(\f_{\hbar}^{-1}\bigr)^{\ast}\tilde f
\ee 
is well defined and belongs to
$C_r^{\ast}((Q\times Q)\times \{\hbar\})$.
The observable $f$ is therefore quantized, $\forall\hbar\leq\hbar(f)$,
by the operator $\Q_{\hbar}(f)\in{\cal K}\bigl(L^2(Q,\mu)\bigr)$:
\be
\label{2.2.10}
\Bigl(\Q_{\hbar}(f)\psi\Bigr)(x)=\int d\mu(x')K^f_{\hbar}(x,x')\psi(x'),\
\psi\in L^2(Q,\mu)\ .
\ee

\section{The torus: a case study}
The 2-torus $T^2$ provides a good test for any quantization scheme. 
Although rather innocent looking, the torus possesses a  set of characteristics 
that make it somewhat special. To begin with, it is not a cotangent bundle, and it is
compact, and therefore the physical expectation is that at the quantum level one will find
only bounded  observables and moreover finite dimensional Hilbert spaces. 
There is, however, another characteristic that distinguishes the torus from
e.g.~the two-sphere, with which it shares the above two properties.
In fact, the Poisson algebra $\cinf(T^2)$ of the torus  does not seem
to admit any subalgebra that separates points  (and contain the constant function 1), 
besides the  algebra $\cinf(T^2)$ itself (and dense subalgebras thereof). 
In particular, it is known that no such finite dimensional 
subalgebra exists \cite{GGG}. 
Thus, from the point of view of canonical quantization, 
it seems that for $T^2$ one is forced to impose the Dirac
 condition (\ref{1.0.11}) on the whole Poisson algebra. But it is also
known that no nontrivial finite dimensional Lie representation of $\cinf(\P)$
can be found, for any connected compact symplectic manifold $\P$ \cite{GiM}. 
On the other hand, any  infinite dimensional  representation of such a 
Poisson algebra will include unbounded operators \cite{Av}.
So, it seems
that every conceivable Dirac-like quantization of the
torus will produce infinite dimensional Hilbert spaces and unbounded observables, 
conflicting with  natural physical expectations.

We discuss next a quantization of $\cinf(T^2)$ proposed by Gotay, which has the great interest
of proving that (irreducible) quantizations of full Poisson algebras can indeed be found.
It does not, however, avoids the above mentioned drawbacks. A modified quantization,
departing from the exact implementation of Dirac's condition, is already presented in the next
section. This presentation follows geometric quantization methods, although at some
point a deformation is introduced. The same quantization is discussed in section \ref{qt},
this time showing that it is directly obtained from a group action such as those
discussed in example 4 of section \ref{gru}.

\subsection{Geometric quantization of the torus}

Let $(\P,\omega)$ be a symplectic manifold and
$\{\,,\}$ the corresponding Poisson bracket.
Let $\S$ be a Lie-subalgebra of $\bigl(\cinf(\P),\{\,,\}\bigr)$,
containing the
constant function 1. By  {\it prequantization} of $\S$ it is meant
a linear map $\Q$ from $\S$ to 
self-adoint operators on a 
 Hilbert space, such that  Dirac's condition
\be
\label{vonN1}
\Q (\{f,g\})=[\Q (f),\Q (g)]_{\hbar }\, \ \ \forall f,g\in\S
\ee
is satisfied and
\be
\label{vonN2}
\Q(1)={\bf 1}\,,
\ee
where {\bf 1} is the identity operator. 

In general,   Dirac's condition 
can be achieved in the full algebra $\cinf(\P)$: given that Hamiltonian vector fields 
provide an (anti-)represen\-tation  of the Poisson algebra,  one could just
adopt the map
$\Q(f)=-i\hbar\xi_f$, where $\xi_f$ is the 
 Hamiltonian vector field defined by $f$, acting on (an appropriate dense domain of) 
  $L^2(\P,\omega^n/n!)$. However, this type of representation 
always leads to
$\Q({1})=0$, which is not acceptable. The formalism of geometric quantization 
(see e.g.~\cite{Wo,We,Pim}) corrects this aspect. In this formalism, the quantization map 
 is of the form
$$
\Q(f)=f-i\hbar\xi_f+\theta(\xi_f)\,,
$$
where $d\theta=\omega$. In general, the 1-form $\theta$ is  defined only locally, 
with $\xi_f+{i\over\hbar}\theta(\xi_f)$ being properly interpreted as a
 covariant derivative on a certain line bundle, which requires  $h^{-1}\omega$
to be of integral  cohomology class. (In this subsection $h$ is Planck's constant.)

Besides (\ref{vonN1}) and (\ref{vonN2}), a true {\em canonical quantization} is required to satisfy
a further set of mathematical physics conditions, one of the most proeminent being irreducibility
(see  \cite{Go} for a thorough discussion). It is this irreducibility condition that typically 
calls for the necessity of a polarization in geometric quantization, leading
to a drastic reduction of the algebra over which the quantization map is defined.

Nevertheless, in \cite{GGT}  (see also \cite{Go}) Gotay shows that a given prequantization of the full algebra
of smooth functions on the torus $T^2$ satisfies even the condition of being irreducible.
Thus, it seems that in this case polarization would not be necessary and that a canonical quantization of
{\em all} smooth observables has been achieved.  
Despite the obvious interest of this result, there is a high price to pay
for this full quantization,  in the sense that  von Neumann's condition 
(\ref{1.0.10})  is  badly broken  (see \cite{mytorus}).
We discuss next Gotay's proposal.

Let $N\in\N$ and  consider the torus $T^2$ of area $Nh$, which we identify  with $\R^2/\Z^2$ equipped with 
the symplectic form
\be
\label{tdois1}
\omega=Nh\,dx\wedge dy\, ,
\ee
where we have introduced local coordinates $(x,y)$.
The  Poisson algebra $\cinf(T^2)$ is therefore the algebra of 
$\cinf$ periodic functions on $\R^2$.
Gotay's quantization, which is obtained through geometric quantization methods, 
can be described as follows. 
Choosing the connection  $\theta=-Nhydx$, the 
 prequantum Hilbert space  $\H_N$ can be seen as the completion of the space ${\cal D}_N$ 
of complex  $\cinf$  functions in
 $\R^2$ such that, 
\be
\label{tdois3}
\phi(x+m,y+n)=e^{2\pi i N n x}\phi(x,y),\ \ \forall m,n\in \Z\,,
\ee
with respect to the inner product
\be
\label{tdois4}
\langle\phi,\phi'\rangle=\int_{[0,1]\times [0,1]} dxdy\,{\bar\phi}\,\phi'\,.
\ee
The prequantization map is given by 
\be
\label{tdois5}
\Q_N(f)\phi=f\phi-{i\over 2\pi N}\left({\partial f\over\partial y}\left({
\partial \phi\over\partial x}-2\pi Niy\phi\right)-{\partial f\over\partial x}
{\partial\phi\over\partial y}\right),\ \ \forall f \in C^{\infty}(T^2)\,.
\ee
Being a prequantization, it is true $\forall N$ that Dirac's condition (\ref{1.0.11}) is fulfilled for all observables in the algebra
$C^{\infty}(T^2)$. The case $N=1$  is special in that  irreducibility conditions are satisfied \cite{GGT}. Thus, it appears that $\Q_1$ above gives a {\em bona fide} canonical quantization
of {\em all} observables in a symplectic manifold.
However, there is no control over the multiplicative structure of the algebra $C^{\infty}(T^2)$,
and therefore there is also no control over the spectrum of the quantum operators $\Q_1(f)$ (\ref{tdois5}). For instance, considering the classical observables 
$\sin(2\pi x)$ and $\cos(2\pi x)$, one can show \cite{mytorus} that 
\be
\label{tdois11}
\Q_1^2\bigl(\sin(2\pi x)\bigr)+\Q_1^2\bigl(\cos(2\pi x)\bigr)  = {\bf 1}+\cal R \, ,
\ee
where $\cal R$ is an {\em unbounded}  operator with no correspondence with any observable.
(The same happens with the functions $\sin(2\pi y)$ and $\cos(2\pi y)$.)
In particular, the spectrum of the quantum operators corresponding to the {\em sinus} and
{\em cosinus} functions is the full line $\R$, and the correlation between the two functions is lost.

\medskip
Let us now discuss a different quantization of the torus $T^2$ which we believe
satisfies appropriate physical requirements.
This quantization can be introduced in a number of ways  (see \cite{ACG,Rif,BMS,Va}). 
Our treatment in this section is inspired in \cite{ACG}.
As we will see in the next section,  the same quantization appears naturally
in the context of noncommutative geometry.

Let us consider then the prequantizations $\Q_N$ (\ref{tdois5}). 
In the geometric   quantization formalism, the way to achieve a quantization starting from  a given prequantization is to restrict the action of observables to (covariantly) constant sections over a given
polarization \cite{Wo,We}. 
Let us then focus on the space of sections $\phi\in\H_N$
such that ${\partial\over\partial y}\phi=0$
  (note that with the connection $\theta=-Nhydx$, the
 covariant derivative $\nabla_y:={\partial\over\partial y}+{i\over\hbar}\theta
\bigl({\partial\over\partial y}\bigr)$ coincides with ${\partial\over\partial 
x}$). Taking (\ref{tdois3}) into account, those sections satisfy 
\be
\label{distoro1}
\bigl(1-e^{2\pi iNx}\bigr)\psi(x,0)=0\,,
\ee
and it is therefore clear that there are no nontrivial solutions in
 $\H_N$. There are, however $N$ independent
 generalized  solutions, of the form
\be
\label{distoro2}
\psi_k=\delta(x-k/N),\ \ k=0,1,\ldots,N-1\,.
\ee
In fact, the distributions\ $\psi_k$ defined by:
\be
\label{distoro3}
\psi_k(\phi)=\int dx\,dy\,\delta(x-k/N)\phi(x,y)=
\int_0^1 dy\,\phi(k/N,y)
\ee
are well defined on the  dense  space ${\cal D}_N\subset\H_N$ and satisfy
\be
\label{distoro4}
\psi_k\Bigl({\partial\phi\over\partial y}\Bigr)=0,\ \ 
\forall\phi\in{\cal D}_N\,.
\ee
(The appearance  of  distributional solutions is common in
geometric quantization; the fibres $x=k/N$ are an example of so-called
 Bohr-Sommerfeld submanifolds \cite{Wo,We}.) Let us then choose the finite dimensional
Hilbert space (isomorphic to  $\C^N$)  generated by the $N$ distributions $\psi_k$, 
with inner product  $\langle\psi_k,\psi_{k'}\rangle=\delta_{kk'}$, to be
the quantum Hilbert  space 
 associated with the 2-torus of area $Nh$.
In general, the choice of a polarization  selects a
restricted subalgebra of observables with a well defined  action on the quantum 
Hilbert space. In the present case, one can easily check that 
a quantum  observable $\Q_N(f)$ is well defined if and only if
\be
\label{distoro6}
\psi_k\Bigl(\Q_N(f){\partial\phi\over\partial y}\Bigr)=\psi_k\Bigl(
\nabla_x\Bigl({\partial^2f\over\partial y^2}\phi\Bigr)\Bigr)=0,\ \ \forall k,\ 
\forall\phi\,.
\ee
It follows that the only functions $f\in\cinf(T^2)$ that are quantized
by this process are the ones which depend exclusively on $x$. These act
simply by
multiplication, i.e.
\be
\label{distoro7}
\Q_N\bigl(f(x)\bigr)\psi_k=f(k/N)\psi_k\,.
\ee
The extension of the  quantization to further observables requires a new look 
at the quantization of  functions  $g(y)\in\cinf(T)$.

Following \cite{ACG}, let us consider the  unitary operators given by $\V_N(b)$, $b\in\R$:
\be
\label{distoro8}
\bigl(\V_N(b)\phi\bigr)(x,y):=\phi(x+b,y),\ \ \phi\in\H_N,
\ee
which are associated with translations  $x\mapsto x+b$.
Clearly, the operators $\V_N(b)$ have a well defined action
in $\H_N$ only for values of $b$ of the form $b=n/N$, $n\in\Z$, 
as follows from  (\ref{tdois3}). These finite translations   correspond
to maps between the Bohr-Sommerfeld leaves, and we therefore get
well defined   unitary 
translation 
operators $\V_N(n)$ on the quantum
 Hilbert space:
\be
\label{distoro9}
\V_N(n)\psi_k:=\psi_{k-n\,({\rm mod}\,N)}\,.
\ee
Taking into account the Weyl-Moyal quantization\footnote{In $T^*\R$, the W-M quantization
(\ref{1.2.5}) of the function $g(p)=e^{ibp/\hbar}$ is in fact the translation operator in
$L^2(\R)$: $\psi(q)\mapsto \psi(q+b)$.} in $T^*\R$, 
one can look at the operators $\V_N(n)$ as the quantization of the functions $e^{2\pi iny}$.
This interpretation is further supported by the commutation relations satisfied by these operators
and the quantization of the functions $e^{2\pi imx}$, namely
\be
\label{distoro11}
\V_N(n)\,\Q_N\bigl(e^{2\pi imx}\bigr)=e^{2\pi imn/N}\Q_N\bigl(e^{2\pi imx}\bigr)\V_N(n)\,.
\ee
Let us then define a quantization map by
\be
\label{distoro12}
\Q_N\Bigl(e^{2\pi i(mx+ny)}\Bigr):=e^{\pi inm/N}\Q_N\Bigl(e^{2\pi imx}\Bigr)\V_N(n).
\ee
Concerning the Dirac rule we obtain in particular 
\ba
\label{distoro13}
\Bigl[\Q_N\bigl(e^{2\pi imx}\bigr)\!\!\!\!\!\! & , & \!\!\!\!\!\! 
\Q_N\bigl(e^{2\pi iny}\bigr)\Bigr]-i\hbar\Q_N\Bigl(\bigl\{e^{2\pi imx},
e^{2\pi iny}\bigr\}\Bigr)= \nonumber \\
& = & \!\!\!\! 2i\Bigl(mn\pi/N-\sin(mn\pi/N)\Bigr)
\Q_N\Bigl(e^{2\pi i(mx+ny)}\Bigr)\,,
\ea
which shows that, for large $N$, the Dirac condition is well approximated by
slow varying functions, i.e., such that the Fourier decomposition contains only components
$e^{2\pi i(mx+ny)}$ of frequencies $m$ and $n$ which are small compared to $N$.
%


\subsection{\cst-algebraic quantization of the torus }
\label{qt}


The Weyl-Moyal quantization map (\ref{1.2.5}), section \ref{WM}, can be immediately rewritten in the form
\be
\label{2.3.3}
\bigl(\Qh(f)\psi\bigr)(x)=\int \left( \int {dp\over 2\pi}\,
e^{-ipy} f\bigl(x+{\hbar \over 2}y,p\bigr)\right)\psi(x+\hbar y)dy\ .
\ee
When comparing (\ref{2.3.3}) with expression (\ref{2.1.5}),
 section \ref{gru},  we see that $\Qh(f)$ coincides with  
$\pi({\hat f_{\hbar}})$, where 
\be
\label{2.3.4}
{\hat f_{\hbar}}(x,y)=\int {dp\over 2\pi}\,
e^{-ipy} f\bigl(x+{\hbar \over 2}y,p\bigr)\, .
\ee
Here,  ${\hat f_{\hbar}}$ should be 
considered as an element of 
$\R >\!{\lhd}_{\alpha ^{\hbar}}\R$.
So, the quantum algebra  coincides precisely with
$C_r^{\ast}(\R >\!{\lhd}_{\alpha ^{\hbar}}\R)$, and the quantization map is given
by $\Qh=\pi\circ\wedge_{\hbar}$, where $\wedge_{\hbar}(f)={\hat f_{\hbar}}$.

In this perspective, the crucial step in the quantization process  consists in
the introduction  of the  algebra $C_r^{\ast}(\R >\!{\lhd}_{\alpha ^{\hbar}}\R)$, 
associated with the 
nontrivial action of the  tangent vectors  in  configuration space.

\medskip

Let us consider again the two dimensional torus $T^2$. The identification
$T^2\cong \R^2/\Z^2$ gives also a bijection between the set of continuous functions on the torus
and the set of periodic continuous functions on $\R^2$:
$$f(x+m,y+n)=f(x,y), \ \ \ \ \ \forall \, m,n\in \Z\, .$$
The partial Fourier transform: 
\be
\label{2.3.22}
{\cal F}\, :\, f(x,y)\mapsto\tilde f(x,n)=\int_0^1 
e^{-2\pi iny}f(x,y)\, dy\, ,
\ee
further allows us to pass from functions on the torus to functions
on $S^1\times\Z$, or periodic functions on $\R\times\Z$. 
For definiteness, let us consider as  complete algebra of regular observables the
subspace
$\A(T^2)$ of those functions whose (total)  Fourier transform possesses only  a finite 
number of nonzero  coefficients.  
It follows that the partial Fourier transform $\cal F$ (\ref{2.3.22}) is an isomorphism
between $\A(T^2)$ and the space $\A(S^1\times\Z)$ of finite linear combinations
of the functions $F_{mk}$ in $S^1\times\Z\,$ defined by:
\be
\label{2.3.23}
F_{mk}(x,n):=e^{2\pi imx}\delta_{nk}\ .
\ee
Let us consider the family of actions $\a^{\p}$ of $\Z$ on $S^1$,
parametrized by real numbers $\p\in[0,1[$ and defined by
\be
\label{2.3.24}
\a^{\p}_n(x)=x+n\p\ {\rm mod}\, 1\ ,\ n\in\Z\, .
\ee
With each of these actions, let us associate the semidirect product groupoid
$S^1 >\!{\lhd}_{\alpha ^{\p}}\Z$, as in examples 4 and 4a in section \ref{gru}.
Between two elements $g_1=(x_1,n_1)$ and  $g_2=(x_2,n_2)$ of
$S^1 >\!{\lhd}_{\alpha ^{\p}}\Z$ such that 
$x_2=\a^{\p}_{n_1}(x_1)=x_1+n_1\p\ {\rm mod}\, 1$,
the composition rule is given by $g_1g_2=(x_1,n_1+n_2)$.
The inverse of the  element 
$g=(x,n)$ is $g^{-1}=(\a^{\p}_n(x),-n)=(x+n\p\ {\rm mod}\, 1,-n)$.
Finally, let us note that both {\it r}-fibres and {\it s}-fibres can be identified with $\Z$.
In particular, given an object 
$x_0\in{\rm Obj}(S^1 >\!{\lhd}_{\alpha ^{\p}}\Z)\cong S^1$,
the corresponding {\it r}-fibre is the set
\be
\label{2.3.25}
G^{x_0}=\{(x_0,n),\, n\in\Z\}\, ,
\ee
and  the corresponding {\it s}-fibre is
\be
\label{2.3.26}
G_{x_0}=\{(\a^{\p}_n(x_0),-n),\, n\in\Z\}\, .
\ee
The discrete structure of the  fibres allows us to define the convolution algebra, 
which we now describe. Consider then the space
$\A(S^1\times\Z)$, whose elements we identify with functions 
$F:\R\times\Z\to\C$ such that $F(x+m,n)=F(x,n),\ m\in\Z$. 
The involution  ${}^\ast$ in $\A(S^1\times\Z)$ is defined by
\ba
\label{2.3.27}
F^{\ast}(g)&=&\bar F(g^{-1}),\ {\rm or}\\
\label{2.3.28}
F^{\ast}(x,n)&=&\bar F(x+n\p,-n),\ F\in\A(S^1\times\Z)\, .
\ea
In particular for the basis elements $F_{mk}$ (\ref{2.3.23}) we get
\be
\label{2.3.29}
F^{\ast}_{mk}=e^{2\pi imk\p}F_{-m\, -k}\, .
\ee
The convolution ${\star}$ is defined by
\be
\label{2.3.30}
(F\star G)(x,n)=\sum_{m\in\Z}F(x,m)G(x+m\p,n-m)\, .
\ee
$\A(S^1\times\Z)$ is therefore an involutive algebra with identity, namely the function
 $F_{00}(x,n)=\delta_{n0}$. In particular, the convolution (\ref{2.3.30}) of basis elements
leads to
\be
\label{2.3.31}
F_{mk}\star F_{m'k'}=e^{2\pi im'k\p}F_{m+m'\, k+k'}\, .
\ee
One can easily show that this algebra is generated by the two  elements $F_{10}$ e $F_{01}$.
Furthermore, since $F_{10}$ are $F_{01}$ are unitary
and satisfy the commutation relations
\be
\label{2.3.39}
F_{01}\star F_{10}=e^{2\pi i\p}F_{10}\star F_{01}\ ,
\ee
we conclude that the algebra in question is none other than the well know
{\em  universal  rotation algebra} $\A^{\rm rot}_{\p}$,
parametrized by $\p$, which is precisely
defined as the  $\ast$-algebra
generated by two  elements $u$ and $v$ subject to the relations
$u^{\ast}u=u u^{\ast}=v^{\ast}v=v v^{\ast}=1$ and
$vu=e^{2\pi i\p}uv$ \cite{W-O}.
The {\em rotation  $C^{\ast}$-algebra} $A^{\rm rot}_{\p}$
is by definition the completion of $\A^{\rm rot}_{\p}$ with respect to the  norm
\be
\label{2.3.40}
\|a\|:={\rm sup}\{\|\pi a\|\ :\ \pi\ \hbox{is a  representation of}\ 
\A^{\rm rot}_{\p}\},\ a\in\A^{\rm rot}_{\p}\, ,
\ee
and satisfies the folowing   universality property \cite{W-O}
\begin{theo}
\label{univrot}
Let A be a   $C^{\ast}$-algebra with two elements $u'$, $v'$
satisfying the same relations as the generators $u$, $v$ of $A^{\rm rot}_{\p}$.
Then there exists a morphism $\vf:A^{\rm rot}_{\p}\to A$ such that
$v\mapsto v'$ and $u\mapsto u'$. If $\p$ is irrational then $\vf$ 
is an isomorphism between $A^{\rm rot}_{\p}$ and the smallest closed subalgebra
of $A$ that contains  $u$ and $v$. 
\end{theo}
This result shows  immediately that 
$C^{\ast}_r(S^1 >\!{\lhd}_{\alpha ^{\p}}\Z)$ is isomorphic to 
$A^{\rm rot}_{\p}$ when $\p$ is irrational.
In the  $\p$ rational case, and for an arbitrary algebra $A$, 
the map $\vf$ of the above theorem  is not necessarily 
injective. In the present case, however, injectivity is clearly ensured, 
and therefore  
$C^{\ast}_r(S^1 >\!{\lhd}_{\alpha ^{\p}}\Z)$ is  isomorphic to 
$A^{\rm rot}_{\p}$, $\forall \p$.

The structure of the   rotation algebras $A^{\rm rot}_{\p}$ depends
heavily on the value of  $\p$: for $\p=0$ we recover, as expected, 
the algebra $C(T^2)$ of continuous functions on the torus;
the irrational  $\p$ case is extremely  interesting from the point of view
of noncommutative geometry and has been extensively studied \cite{Rif,Co,Va,An}.
Let us focus on the  rational (nonzero) $\p$ case, following \cite{Va,An}. 
As we will see shortly, the quantization (\ref{distoro12}) described in the previous section 
will emerge here quite naturally.

Let us first note that two distinct values $\p$, $\p'$ such that $\p+\p'=1$
lead to the same algebra, i.e.   $A^{\rm rot}_{1-\p}$ is isomorphic to
 $A^{\rm rot}_{\p}$. 
Let then $\p={K\over N}$,  with $K, N\in\N$ and  $K\leq N/2$.
Taking into account the convolution
(\ref{2.3.31}), one can easily check that the elements of the  form $F_{mN\, kN}$,
$m,k\in\Z$, commute with the generators, and therefore belong to the centre of the algebra.
So, given any irreducible representation  $\pi^0$, the image $\pi^0(F_{mN\, kN})$ of those elements
must be proportional to the identity.
It follows that the   irreducible representations of the algebra $A^{\rm rot}_{\p}$,
with $\p=K/N$ as above, are finite dimensional, of dimension $N$.  Note also
that, being finite dimensional, the irreducible representation 
is unique (modulo unitary equivalence).

A convenient irreducible representation can be easily found, as follows.
Let   ${\cal H}_N\cong \C^N$ be the  Hilbert space generated by an orthonormal set
of $N$ vectors, say
$\{v^0,v^1,\ldots,v^{N-1}\}$. In $B({\cal H}_N)$ consider 
unitary operators $\U_N(K)$, $\V_N(K)$ such that
\be
\label{2.3.45}
\U_N(K)v^k=e^{2\pi ik/N}v^k\, ,
\ee
\be
\label{2.3.46}
\V_N(K)v^k=v^{k-K\, \hbox{mod}\, N} .
\ee
We obtain immediately the commutation relations: 
\be
\label{2.3.50}
\V_N(K)\U_N(K)=
e^{2\pi iK/N}\U_N(K)\V_N(K)\, .
\ee
The pair $\U_N(K)$, $\V_N(K)$ therefore satisfies the  relations  (\ref{2.3.39}) 
corresponding to $\p=K/N$, which shows that the $\ast$-morphism
$\pi_{K,N}: A^{\rm rot}_{\p}\to B({\cal H}_N)$ given by
\ba
\label{2.3.51}
\pi_{K,N}(F_{10})&=&\U_N(K)\\
\label{2.3.52}
\pi_{K,N}(F_{01})&=&\V_N(K)
\ea
is well defined and is a representation, obviously  
irreducible, of $A^{\rm rot}_{K/N}$.

Let us finally construct a family $\Q_{K/N}$ of quantizations of the 2-torus.
As already suggested, let us adopt as complete algebra of regular observables
the subalgebra
$\A(T^2)\subset C(T^2)$ of those functions whose Fourier transform
possesses only a  finite number of nonzero coefficients.
Let then
$\wedge_{K/N}: \A(T^2)\to C^{\ast}_r(S^1 >\!{\lhd}_{\alpha ^{\p}}\Z)$
denote the maps given by
\be
\label{2.3.53}
f(x,y)\mapsto {\hat f}_{K/N}(x,n)=
\int_0^1 
e^{-2\pi iny}f(x+nK/2N,y)\, dy\, ,
\ee
which correspond to (\ref{2.3.4}).
For the elements  $e^{2\pi i(mx+ky)}$ of the base we get simply
\be
\label{2.3.54}
e^{2\pi i(mx+ky)}\stackrel{\wedge_{K/N}}\longmapsto e^{\pi imkK/N}F_{mk}\ .
\ee
The quantizations maps  are then 
$\Q_{K/N}=\pi_{K,N}\circ \wedge_{K/N}$, leading to
\be
\label{2.3.55} 
\Q_{K/N}\bigl(e^{2\pi i(mx+ky)}\bigr)=
e^{\pi imkK/N} \U_N(K)^m\V_N(K)^k\in B({\cal H}_N).
\ee
In particular for $K=1$, this coincides with the quantization put forward in the previous section,
expressed namely in quantization rules (\ref{distoro7}), (\ref{distoro9})
and (\ref{distoro12}).



\begin{thebibliography}{99}

\bibitem{HRS}R. Honegger, A. Rieckers, L. Schlafer, {\em SIGMA} {\bf 4} (2008) 047

\bibitem{W} S. Waldmann, {\em J. Geom. Phys.}  {\bf 81} (2014) 10



\bibitem{FR} K. Fredenhagen, K. Rejzner, {\em QFT on curved spacetimes:
axiomatic framework and examples}, arXiv:1412.5125 [math-ph]


\bibitem{ST} A. Stottmeister, T. Thiemann, 
{\em Coherent states, quantum gravity and the Born-Oppenheimer approximation, III: Applications to loop quantum gravity}, arXiv:1504.02171 [math-ph]



\bibitem{Fo} G. B. Folland, {\it Harmonic Analysis in Phase Space}
(Princeton University Press, 1989)

\bibitem{Lan} N. P. Landsman, {\it Mathematical Topics between Classical
and Quantum Mechanics} (Springer-Verlag, New York, 1998)

\bibitem{Va} J. C. V\'arilly, {\it An Introduction to Noncommutative
Geometry} (EMS Series of Lectures in Mathematics, 2006) 

\bibitem{SW} A. Cannas da Silva, A. Weinstein, {\em Geometric Models for Noncommutative Algebras} 
(Berkeley Mathematics Lecture Notes series, AMS, 1999)

\bibitem{RL} R. Loja Fernandes, {\em Deformation Quantization and Poisson Geometry}, Resenhas IME-USP {\bf 4} (2000) 327
 

\bibitem{Co} A. Connes, {\it Noncommutative Geometry} (Academic Press, 
London, 1994)

\bibitem{Dir} P. Dirac, {\it The Principles of Quantum Mechanics}
(Oxford University Press, 4th ed. 1967)

\bibitem{Go} M. J. Gotay, {\it Obstructions to Quantization},
in Mechanics: From Theory to Computation
(Essays in Honour of Juan-Carlos Simo), J. Nonlinear Sci. Eds. (Springer, New
York, 2000) 171

\bibitem{Ri1} M. A. Rieffel, {\it Deformation Quantization for Actions 
of $\R^d$}, Memoirs Am. Math. Soc. {\bf 106}, Nr. 506 (1993) 

\bibitem{B-R} O. Bratteli, D. W. Robinson, {\it Operator Algebras and
Quantum Statistical Mechanics 1} (Springer Verlag, New York, 1987)

\bibitem{R-S} M. Reed, B. Simon, {\it Methods of Modern Mathematical
Physics}, vol. I (Academic Press, 1980)








\bibitem{Lan1} N. P. Landsman, {\em J. Geom. Phys.} {\bf 12}  (1993) 93



\bibitem{VaC} J. F. Cari\~nena, J. Clemente-Gallardo, E. Follana, 
J. M. Gracia-Bond\'\i a, A. Rivero, J.~C. V\'arilly, {\em J. Geom. Phys.} {\bf 32}  (1999) 79

\bibitem{GGG} M. J. Gotay, J. Grabowski, H. B. Grundling, {\em Proc. Amer.
Math. Soc.} {\bf 128}  (2000) 237

\bibitem{GiM} V. L. Ginzburg, R. Montgomery, {\it Geometric Quantization
and No Go Theorems}, Banach Center Publications {\bf 51} (2000) 69



\bibitem{Av} A. Avez, {\em C. R. Acad. Sci. Paris} {\bf A279}  (1974) 785

\bibitem{Wo} N. M. J. Woodhouse, {\it Geometric Quantization} (Clarendon
Press, Oxford, 1992)


\bibitem{We} A. Weinstein, {\it Deformation Quantization}, S\'em.
Bourbaki {\bf 789}, Ast\'erisque {\bf 227} (1995) 389



\bibitem{Pim} J. P. Nunes, {\em Rev. Math. Phys.} {\bf 26}   (2014) 1430009


\bibitem{GGT} M. J. Gotay, {\it On a Full Quantization of the Torus}, 
in Quantization, Coherent States and 
Complex Structures, eds. J.-P. Antoine {\it et al.} (Plenum, New York, 1995) 

\bibitem{mytorus} J. M. Velhinho, {\em Int. J. Mod. Phys. A} {\bf 22} (1998) 3905

\bibitem{ACG} V. Aldaya, M. Calixto, J. Guerrero, {\em Comm. Math. Phys.}
{\bf 178} (1996) 399

\bibitem{Rif} M. A. Rieffel, {\em Comm. Math. Phys.} {\bf 122} (1989) 531

\bibitem{BMS} M. Bordemann, E. Meinrenken, M. Schlichenmaier,
{\em Comm. Math. Phys.} {\bf 165} (1994) 281


\bibitem{W-O} N.E. Wegge-Olsen, {\em K-Theory and C*-algebras - a friendly approach} (Oxford University Press, 1993) 

\bibitem{An} N. F. Ant\'onio, {\it \'Algebras-$C^*$ de Rota\c c\~ao : 
Propriedades Elementares e de Estrutura}, Disserta\c c\~ao de Mestrado 
(IST-UTL, Lisboa, 1998)








\end{thebibliography}
\end{document}